\documentclass[fleqn,usenatbib]{mnras}
\usepackage{hyperref}
\usepackage{newtxtext,newtxmath}
\usepackage[T1]{fontenc}

\DeclareRobustCommand{\VAN}[3]{#2}
\let\VANthebibliography\thebibliography
\def\thebibliography{\DeclareRobustCommand{\VAN}[3]{##3}\VANthebibliography}

\usepackage{graphicx}	% Including figure files
\usepackage{amsmath}	% Advanced maths commands

\title[JWST Variability Modelling of LSR~J1835+3259]{JWST/NIRISS-SOSS Spectroscopic Variability and Auroral Modelling of
Warm Ultracool Dwarf LSR J1835+3259}

\author[E. Pike et al.]{
    Eleanor Pike,$^{1}$\thanks{E-mail: efpike1@sheffield.ac.uk}
    Stuart Littlefair,$^{1}$
    J. Sebastian Pineda, $^{2}$
    Melodie Kao, $^{3}$
    Fei Wang, $^{4}$
    Ben Burningham, $^{4}$
    \newauthor
    David García-Álvarez, $^{5,6}$
    Gregg Hallinan, $^{7}$
    Caroline Morley, $^{8}$
    John E. Gizis, $^{9}$
    John D. Nichols$^{10}$
\\
$^{1}$Astrophysics Research Cluster, School of Mathematical and Physical Sciences, University of Sheffield, Sheffield, S3 7RH, UK\\
$^{2}$Laboratory for Atmospheric and Space Physics, 3665 Discovery Drive Boulder, CO 80303, USA\\
$^{3}$Lowell Observatory, 1400 W Mars Hill Road, Flagstaff, AZ 86001, USA\\
$^{4}$Department of Physics, Astronomy and Mathematics, University of Hertfordshire, AL10 9AB, Hatfield, UK\\
$^{7}$Instituto de Astrofísica de Canarias, Avenida Vía Láctea, 38205 La Laguna, Tenerife, Spain \\
$^{8}$Grantecan S. A., Centro de Astrofísica de La Palma, Cuesta de San José, 38712 Breña Baja, La Palma, Spain\\
$^{5}$Cahill Center for Astronomy and Astrophysics, MC 249-17, California Institute of Technology, Pasadena, CA 91125, USA\\
$^{6}$Department of Astronomy, University of Texas at Austin, 2515 Speedway, Stop C1400, Austin, TX 78751, USA\\
$^{9}$University of Delaware, Data Science Institute, Newark, DE 19716, USA\\
$^{10}$School of Physics and Astronomy, University of Leicester, Leicester, UK}

\date{Accepted XXX. Received YYY; in original form ZZZ}

\pubyear{\the\year{}}

\begin{document}
\label{firstpage}
\pagerange{\pageref{firstpage}--\pageref{lastpage}}
\maketitle

\begin{abstract}
Auroral variability in ultracool dwarfs (UCDs)  has been proposed as a key source of optical and infrared modulation, yet its physical origin remains poorly understood. We present \textit{JWST} Near Infrared Imager and Slitless Spectrograph/Single Object Slitless Spectroscopy (NIRISS/SOSS) time-series spectroscopy of the M8.5 UCD LSR J1835+3259.  These observations probe the object's wavelength-dependent rotational variability, providing insight into the atmospheric processes underlying its infrared behaviour. Additional GTC/MEGARA optical spectra reveal that its H$\alpha$ emission has declined markedly since earlier epochs, suggesting a reduced activity state during the \textit{JWST} observations. We detect no emission from H$_3^+$ or absorption by hydronium in the NIRISS spectra. We use principal component analysis to identify two distinct modes of spectral variability: one dominated by FeH modulation near 1 $\mu$m and another contributing broadband  variability across the H and K bands. We find that no model atmosphere grid provides a reasonable fit to the average NIRISS spectrum. However, a blended model comprising an M-dwarf photosphere and a cooler blackbody component provides an improved fit to the mean spectrum, the broadband spectral energy distribution and the NIRISS broadband variability. By adding perturbations to PICASO model grids, we demonstrate that a combination of atmospheric heating and non-equilibrium chemistry can successfully mimic a blend of unperturbed photosphere and a blackbody, and that the presence of these effects may explain the unusual spectrum and variability of LSR~J1835+3259.

\end{abstract}

\begin{keywords}
brown dwarfs -- planets and satellites: atmospheres -- stars: atmospheres -- planets and satellites: aurorae
\end{keywords}

%%%%%%%%%%%%%%%%%%%%%%%%%%%%%%%%%%%%%%%%%%%%%%%%%%

%%%%%%%%%%%%%%%%% BODY OF PAPER %%%%%%%%%%%%%%%%%%

\section{Introduction} \label{sec: Intro}

Ultracool dwarfs (UCDs) are canonically defined as stellar/substellar objects of spectral type M7 and later~\citep{1997_Kirk}, forming the bridge between the stellar and planetary regimes. Since many are free-floating, isolated objects, they give the unique opportunity to study giant exoplanet-analogue atmospheres without contamination from stellar hosts.

Radio emission is observed in a substantial fraction of UCDs, with \cite{2024Kao} estimating a quiescent radio occurrence rate of 15--20 per cent across the UCD population and 17--25 per cent for M dwarfs. This radio emission is seen across the entire UCD regime, irrespective of effective temperature, and can be split into two components. The first component consists of quiescent emission that is quasi-stable and nonthermal. It is often attributed to optically thin, relativistic gyro-magnetic emission: radio emission due to the spiralling of relativistic electrons through the magnetic field~\citep{2023Climent}. It was suggested by~\cite{2006Hallinan} that this emission traces out regions densely populated by highly energetic particles trapped within the UCD’s magnetosphere, an analogue to the Jovian radiation belt. Resolved imaging of M8.5 UCD LSR~J1835+3259 (LSR~J1835) has since provided corroborating evidence for this theory \citep{2023Kao, 2023Climent}. The second component consists of highly polarised, coherent radio pulses, consistent with electron-cyclotron maser instability (ECMI) emission~\citep{2007Hallinan}. This same mechanism is responsible for the auroral radio emissions observed from magnetised planets in the Solar System (e.g. \citealt{Treumann_2006, Zarker_1998}).

On the same period as these radio pulses, several UCDs have also been observed exhibiting optical and IR variability~\citep{2013Harding}. \cite{Hallinan_2015} hence proposed a single mechanism to explain the periodically varying emission: an electron beam, which generates the radio emission, interacting with the upper atmosphere of the UCD, thereby creating the auroral surface feature responsible for the optical/IR variability. This mechanism is analogous to Jupiter’s auroral system, further emphasising the suitability of UCDs as massive Jovian analogues. 

While auroral activity is expected to drive some optical/IR variability, it is not the only driver as optical/IR variability is also common amongst non-radio emitting UCDs (e.g. \citealt{2001Bailer-JonesandMundt, 2015Metchev, 2022Vos}). Several alternative mechanisms have been proposed, including inhomogeneous silicate cloud/dust cover~\citep[e.g.][]{2014Radigan, 2023Vos}, thermochemical instabilities~\citep{2016Tremblin}, and upper atmospheric hot spots~\citep{2014Robinson&Marley, 2014Morley}. Since multiple drivers of variability are likely to be acting simultaneously, disentangling them can be challenging. Multi-wavelength observations can help distinguish between different atmospheric processes because different wavelengths probe different atmospheric depths and respond differently to changes in temperature, cloud opacity, and atmospheric composition \citep[e.g.][]{2012Buenzli, 2026OToole}. Comparing the wavelength dependence of the variability can therefore provide constraints on its physical origin. In this regard, \textit{JWST} offers invaluable opportunities, with its exquisite precision and unprecedented wavelength range not achievable via ground-based observations. 

To date, \textit{JWST} has provided evidence for IR auroral activity on two UCDs: CWISEP J193518.59-154620.3 ($\gtrsim\rm Y1$)~\citep{2024Faherty} and SIMP J013656.5+093347.3 (T2.5)~\citep{2025McCarthy, 2025Nasedkin}, the latter having previously also shown evidence for radio aurorae~\citep{2016Kao, 2018Kao}. Atmosphere retrievals for both objects indicate the presence of a $\sim$ 300~K temperature inversion in the upper atmosphere ($\sim$1-10 mbar). These inversions have been attributed to auroral heating via electron precipitation, similar to that observed in Jupiter's upper atmosphere (e.g. \citealt{2025_jupiter_heat}).

Both these objects, however, are towards the cooler end of the UCD regime. To therefore create a unified theory of UCD aurorae, it is vital we investigate auroral activity across a range of spectral types and electron beam energies. This motivates the study of warmer UCDs such as LSR~J1835. LSR~J1835 is an M8.5 UCD, placing it at the boundary between low-mass stellar objects and warm brown dwarfs. Located 5.69~pc away~\citep{GaiaDr3}, it has a rapid rotation period of approximately 2.84 hours, independently measured in both radio~\citep{Hallinan2008} and optical~\citep{2013Harding} observations. The surface gravity, and hence the mass and age, is uncertain. Based on modelling the Na I and TiO line profiles,~\cite{Berdyugina_2017} claim a low surface gravity $\log g=4.50\pm0.05$ and a young age.  In contrast~\cite{2015Filippazzo} derive a substantially higher surface gravity of $\log g=5.22\pm0.11$ using a semi-empirical approach based on the bolometric luminosity and evolutionary models, consistent with an older field-dwarf age. This is discussed further in Section~\ref{section average}. LSR J1835 was the first extrasolar object to show evidence for auroral emission~\citep{Hallinan_2015} and a radiation belt~\citep{2023Kao}, making it a benchmark warm UCD through which to understand how auroral activity may induce optical/IR variability. 

Previously, \cite{Hallinan_2015} modelled the variations in optical Keck/LRIS spectra over LSR~J1835's rotation period via a blended mix of an M-dwarf and blackbody spectrum, the latter representing the emission from a surface feature. They argued that the auroral electron beam’s impact on the upper atmosphere would liberate electrons, leading to an increased density of negative hydrogen ions, creating an optically thick region. An alternative to the explanation proposed by \cite{Hallinan_2015} is that the auroral heating could produce an isothermal region or temperature inversion within the upper atmosphere, with the emergent flux from these pressures similarly producing a blackbody spectrum. 

In this paper, we present \textit{JWST}/Near Infrared Imager and Slitless Spectrograph (NIRISS) time-series spectroscopy of LSR~J1835 aimed at testing these competing explanations for its potential auroral variability. The broad 0.85--2.83~$\mu$m wavelength coverage of NIRISS enables us to characterise the wavelength dependence of the variability and test proposed mechanisms involving H$^-$ opacity, whose contribution changes significantly across this wavelength range. We also present Gran Telescopio Canarias (GTC)/Multi-Espectr\'ografo en GTC de Alta Resoluci\'on para Astronom\'ia (MEGARA) observations tracking the optical H$\alpha$ emission. Obtained two days after the NIRISS observations, these provide an indication of the activity state of LSR~J1835 close to the \textit{JWST} epoch. We also analyse archival 2016 KECK/High Resolution Echelle Spectrometer (HIRES) time-series spectra to establish a longer baseline of H$\alpha$ observations.

Section~\ref{sec: observe and reduce} describes the acquisition and reduction of the MEGARA, HIRES and NIRISS datasets. In Section~\ref{sec: Results}, we first discuss the H$\alpha$ emission and its implications for the activity state of LSR~J1835. We then outline our analysis of the reduced NIRISS spectra, beginning with a principal component analysis of the wavelength-dependent variability, followed by forward-modelling with model atmosphere grids to fit the average spectrum. We report a search for H$_3^+$ and hydronium, present fits using two-column atmosphere models, and explore perturbation experiments altering temperature and chemical abundances. We summarise our main findings in Section~\ref{sec: conclusions}.

\section{Observations and Data Reduction} \label{sec: observe and reduce}

\subsection{GTC/MEGARA} \label{sec: Megara}

The observations of LSR J1835's H$\alpha$ emission were carried out with MEGARA (Gil de Paz et al. 2018; Carrasco et al. 2018) at the GTC under program 26-MULTIPLE-2/22B (PI: D. García-Alvarez). MEGARA is an optical fibre-fed spectrograph mounted on the GTC, a 10.4m telescope in La Palma (Canary Islands, Spain). We used the instrument in the Integral Field Unit (IFU) spectroscopic mode. The IFU  provides a field of view (FOV) of 12.5$\arcsec~\times$~11.3$\arcsec$ and includes eight additional fibre bundles located at the edge of the FOV to enable simultaneous sky subtraction. The spatial sampling is 0.62$\arcsec$ per fibre\footnote{This size corresponds to the diameter of the circle on which the hexagonal spaxel is inscribed}. We selected the HR-R Volume Phase Holographic grating, with a resolution of R $\sim$ 18, 700 and spectral range coverage 640.56--679.71 nm. 

We obtained 55 x 500 s spectra for our observations of LSR~J1835. We observed the standard star HR8634 with the same configuration as the scientific target. The observations took place from 21:30UT until 04:30UT on July 13th of 2023. The observing conditions were clear skies and seeing values of about 1.0--1.2~$\arcsec$ until 03:00UT. From 03:00UT until 04:30UT seeing was <1.5~$\arcsec$, with high thin clouds present.

The raw IFU spectra were reduced using the \texttt{MEGARA DRP} Python package~\citep{2023Pascual, 2025pascual} (v0.13.0), following the procedures outlined in the MEGARA Reduction Cookbook~\citep{cookbook}. Bias subtraction was performed using calibration frames obtained during the same observing night. Defective pixels were automatically masked, and cosmic rays were removed. The spectra were then traced using halogen lamp exposures to identify the fibre positions. Wavelength calibration was carried out using ThNe arc lamp images. A fibre-flat correction was applied using the same halogen lamp exposures. For each observation, the calibrated exposures were combined using a median, which also enabled additional rejection of cosmic rays. The sky background was estimated from dedicated sky fibres and subtracted to produce the final reduced spectra. To extract the stellar signal, we selected all spaxels where the source was clearly detected above the noise. Given the seeing conditions during our observations (1.0--1.5~$\arcsec$), this typically corresponded to the central spaxel and one or two surrounding hexagonal rings, centred on the brightest spaxel.

\subsection{KECK/HIRES}

LSR~J1835+3259 was observed on the night of 2016, July~09 (UT) for a total of $\sim$8.5 hr using the Keck telescopes on Mauna Kea with HIRES ( PI: J. S. Pineda). HIRES was in the red configuration with a 1.148~$\arcsec$  wide slit providing a resolution of $\sim$ 37,000 \citep{Vogt1994}. The echelle and cross disperser angles were set to $-0.235$ and 1.736, respectively, enabling spectral coverage from 570 nm~to 1000 nm~across the 3 different CCD chips of HIRES. The detector gain was set to low and the pixels binned 2x1, in the spatial versus the spectral directions. We also used the order blocking filter OG~530. Seeing was $\sim$ 0.75~$\arcsec$  for the majority of the night with individual exposures lasting 600 s.

We reduced the spectra using the MAKEE pipeline for HIRES.\footnote{See \href{http://www.astro.caltech.edu/~tb/ipac_staff/tab/makee/index.html}{http://www.astro.caltech.edu/$\sim$tb/ipac\textunderscore staff/tab/makee/index.html}} The pipeline bias subtracts the data, averages together the flat field calibration images taken at the beginning and end of the night, divides each data exposure by the combined flat field, and computes a wavelength solution from ThAr arc lamp exposures taken before and after the observing run. MAKEE then extracts the spectra for each order of the dispersed spectrum, subtracting the nearby sky emission. 

\subsection{\textit{JWST}/NIRISS}

\begin{figure*}
 \centering
 \includegraphics[width=\textwidth]{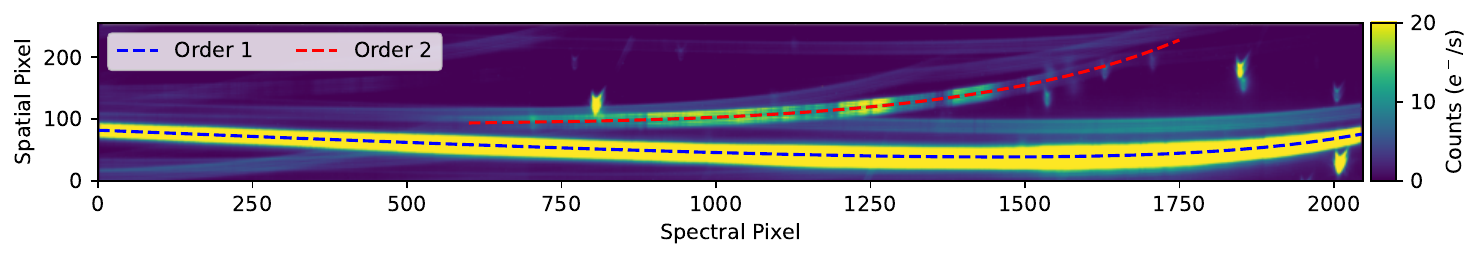}
 \caption{Visualisation of calibrated NIRISS data frames overlaid with the centroid traces of orders 1 and 2. Note this is prior to the \texttt{jwst} \texttt{PhotomStep}, hence the units are in count rates, and that an upper limit of 20~$e^-/$s has been imposed to better show the contaminant orders from field stars. The order 2 trace is so heavily contaminated that we are unable to consider it within our analysis.}
 \label{fig:contamination}
\end{figure*}

 LSR~J1835 was observed using the single object spectroscopy mode (SOSS) of \textit{JWST}'s NIRISS instrument {\bf as part of programme 1874}. Observations started on UTC 2023 July 11 and lasted for approximately 3.6 hours, covering the UCD's full rotation period.  It used the standard GR700XD grism/CLEAR filter combination and subarray SUBSTRIP256 (2048 × 256 pixels), which covers all 3 orders. There was a total of 475 integrations, with 4 groups per integration. A secondary observation was also taken using the F277W filter instead of the CLEAR for a total of 30 integrations, which limits wavelengths to $\lambda >$ 2.6~$\mu$m~\citep{supreme-spoon}.

To reduce the GR700XD/CLEAR dataset, we used a pipeline based on the \texttt{exoTEDRF} (EXOplanet Transit and Eclipse Data Reduction Framework) pipeline (v 1.1.7)~\footnote{Formerly known as \texttt{supreme-SPOON}.}, which in turn is based upon the official STScI \texttt{jwst} calibration pipeline~\footnote{\url{https://jwst-pipeline.readthedocs.io/en/latest/jwst/pipeline/index.html}. \texttt{jwst} v 1.12.5 was used in this work.}. For stage 1, which corrects for errors at the detector level, we followed the procedures outlined by \cite{supreme-spoon} closely. This includes applying \texttt{exoTEDRF}'s custom 1/f noise correction at the group level and omission of the dark current subtraction, which was found to reintroduce said 1/f noise~\citep{supreme-spoon}. 

Stage 2, consisting of instrumental and SOSS model calibrations, was also kept largely the same. However, we diverged in the \texttt{TracingStep}, which determines the centroids of different order traces. We found that  \texttt{APPLESOSS}'s \citep{applesoss} edgetrigger algorithm  originally used in this step struggled to provide reasonable centroids for order 2. This was due to the high level of contamination from other nearby spectral orders of field stars, as visible in Figure~\ref{fig:contamination}. Instead, we used the \texttt{PASTASOSS} package to determine the centroids. Order 1 and 2 traces have pivot points that the traces rotate around for each different observation~\citep{pastasoss_centroids}. The rotation angles around these pivot points are linearly correlated to the pupil wheel position angle of the observation. To therefore calculate the centroids, \texttt{PASTASOSS} uses the pupil wheel position angle to apply rotation transformations to the model traces, giving subpixel accurate positions of the centroids~\citep{pastasoss_centroids}. \texttt{exoTEDRF} has since implemented the use of \texttt{PASTASOSS} in this step in later versions.

In stage 3, we first converted from count rates to units of flux density by applying the \texttt{jwst} \texttt{PhotomStep}. This step is not included in  \texttt{exoTEDRF}  as it is optimised for exoplanet atmospheric spectroscopy, which relies on relative measurements ~\citep{supreme-spoon}. Finally, we performed 1D spectral extraction. We used simple box aperture extraction over the ATOCA (Algorithm to Treat Order ContAmi-
nation)~\citep{2022Atoca} method, as we found the latter introduced wavelength shifts to our spectra. Again, this was due to contamination from spectral traces of nearby field stars. We chose an aperture width of 30 pixels.

F277W filter exposures were reduced similarly to the GR700XD/CLEAR data. While these exposures can be used to identify and correct for contaminant order 0s (see~\citealt{supreme-spoon}), our GR700XD/CLEAR spectra are mainly contaminated by higher order spectra. We therefore found that this made a negligible improvement to the data. Due to these higher order contaminants and poor signal-to-noise ratio (SNR), the resulting order 2 spectra were determined to be unusable. We therefore proceed with only order 1.

\section{Results} \label{sec: Results}
\subsection{Variability in H$\alpha$} \label{sec:Halpha}
\begin{figure}
 \centering
 \includegraphics[width=\columnwidth]{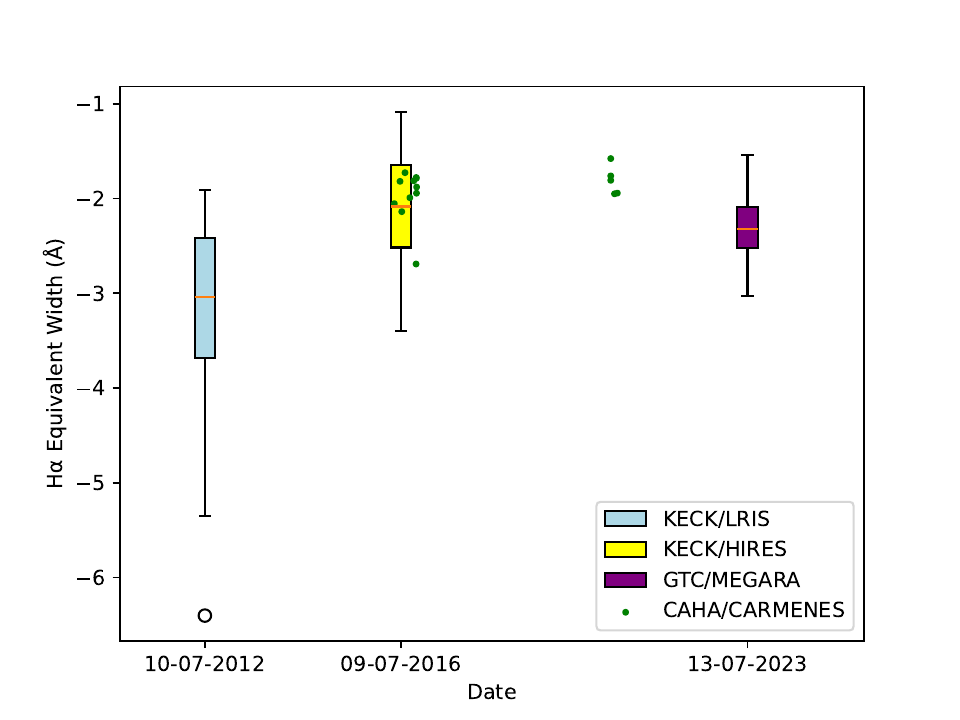}
 \caption{H$\alpha$ equivalent width vs time. The H$\alpha$ EWs vary strongly over a rotation period, leading to a significant spread at any epoch. The box and whisker plots represent data from three instruments: LRIS, HIRES and MEGARA, with the filled region showing the interquartile range, the orange centre line indicating the median, and the endpoints the total range. Outliers are represented by unfilled scatter points that extend beyond 1.5 times the interquartile range. The CARMENES dataset is represented separately via scatter points, as these observations do not cover a full rotation period and instead were taken at several different epochs. The MEGARA data was taken two days after the NIRISS observations.}  
 \label{fig:halpha}
\end{figure}

H$\alpha$ is commonly used as a tracer of magnetic and auroral activity for UCDs ( e.g. \citealt{2008West, 2015Schmidt, 2017Paulo, 2017Pineda}). In particular, H$\alpha$ emission has been observed in association with auroral radio emission and shown to correlate with radio auroral activity \citep{Hallinan_2015, 2017Pineda}. We therefore use the MEGARA dataset to calculate the H$\alpha$ equivalent widths (EWs) as an indicator of LSR~J1835's activity at the time of the NIRISS observations. We define the spectral feature window as the flux contained within 6559.6--6569.6~\AA. To estimate the continuum, we take the average of the regions to the left and right of the feature: 6550--6555 \AA~and 6576--6581 \AA, and then average them. We use the convention of negative EWs indicating emission.

To place the MEGARA H$\alpha$ measurements in context, we compare to previous observations of LSR~J1835: KECK/LRIS on UTC July 10th 2012~\citep{Hallinan_2015} and KECK/HIRES July 9th 2016, which both covered over 1 rotation period, and CAHA/CARMENES with multiple observations from 2016 -- 2020~\citep{Carmenes}. The LRIS EWs were calculated using the same method as the MEGARA data. EWs in the HIRES data set were determined by adding up the flux between 6560~\AA~and 6565~\AA~and normalising by the median of the flux in windows from 6545--6555 \AA~and 6572--6580 \AA~\citep{Reiners2010}. The CARMENES EWs were pre-calculated. The EWs are shown in Figure~\ref{fig:halpha}, with LRIS, HIRES, and MEGARA presented as box-and-whisker plots to illustrate the significant variability in EW over a rotation period as active regions rotate in/out of view.

The LRIS epoch exhibits the strongest median H$\alpha$ emission ($\sim-3$ Å) and the largest variability of any dataset. After 2012, the strength of the H$\alpha$ emission remained at a relatively consistent lower level across the 2016 HIRES, the multiple CARMENES epochs spanning 2016--2020, and the 2023 MEGARA epoch. While the H$\alpha$ strength has changed little between these later epochs, the amplitude of the variability has decreased from the HIRES to the MEGARA epoch.

While we lack simultaneous radio observations and therefore cannot definitively determine the instantaneous auroral activity state during the NIRISS observations, the persistence of weaker H$\alpha$ emission across all post-2012 datasets, together with the progressive reduction in its variability, suggests that the 2023 MEGARA observations represent part of a longer-term lower H$\alpha$ activity state. We therefore consider it plausible that the NIRISS observations sampled the same weakened activity state, although short-term variations between the NIRISS and MEGARA observations cannot be excluded. This interpretation provides a possible explanation for the reduced variability observed in the NIRISS data (see Section~\ref{sec: Dynamic spectrum/PCA}), given the stronger near-IR variability observed by \cite{Hallinan_2015} during the more active 2012 epoch. Although the wavelength coverage does not fully overlap, this suggests that the low near-IR variability observed by NIRISS may reflect a lower activity state rather than intrinsically weaker near-IR variability. Simultaneous H$\alpha$, radio, and near-IR observations would provide a more direct assessment of the activity state during future observations of LSR~J1835.

\subsection{Average NIRISS Spectrum} \label{section average}

\begin{figure*}
 \centering
 \includegraphics[width=\textwidth]{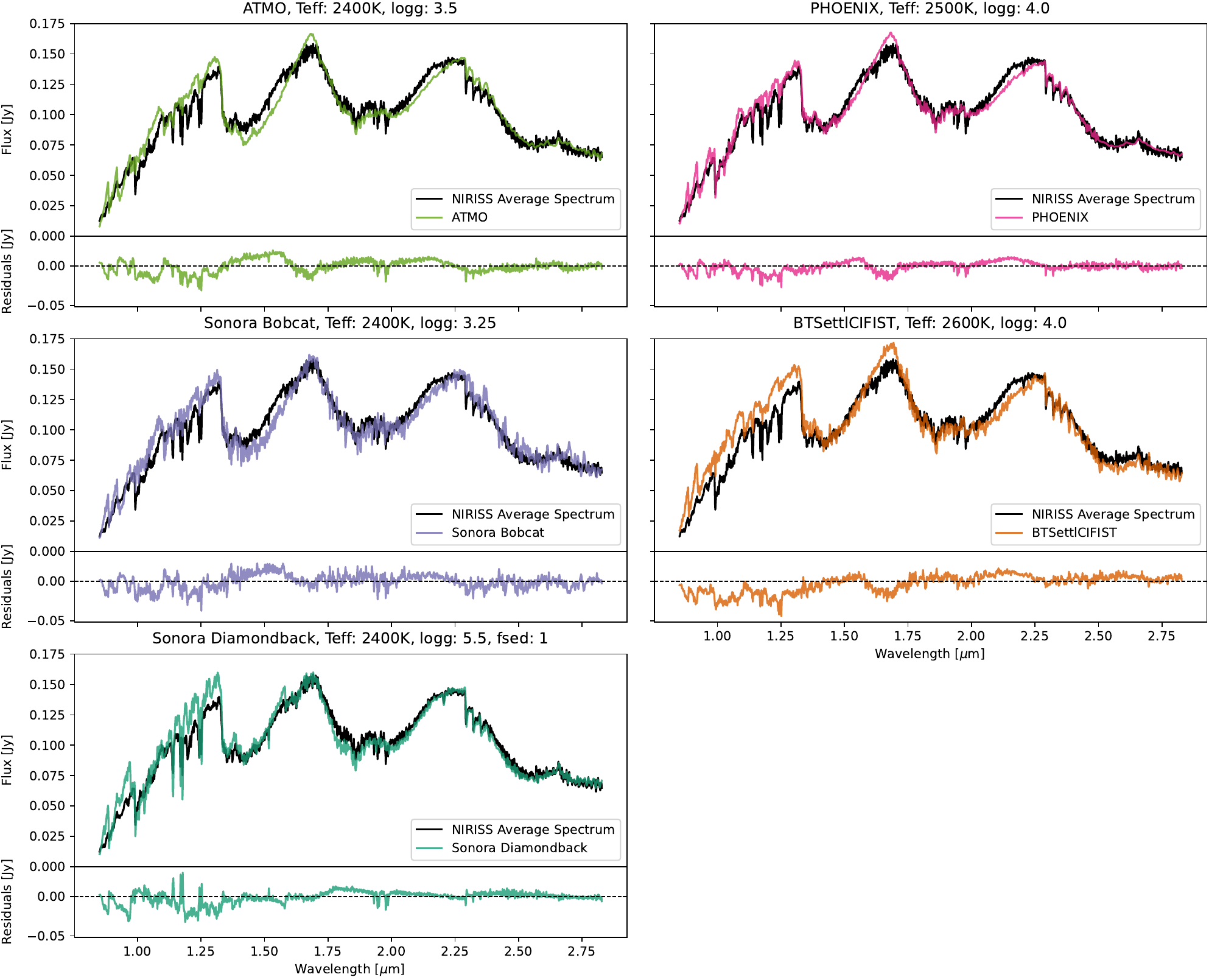}
 \caption{Time-averaged spectrum for LSR~J1835 (black) compared to the best fit model spectra from \texttt{ATMO}, \texttt{PHOENIX}, \texttt{Sonora Bobcat}, \texttt{BTSettl(CIFIST)} and \texttt{Sonora Diamondback} model grids with their associated parameters. The corresponding residuals (average spectrum $-$ model atmosphere spectrum) are shown beneath each model fit. None of the model grids are able to produce a satisfactory fit to the average spectrum, indicating a blended combination of spectra is required.}
 \label{fig:all}
\end{figure*}

The time-averaged NIRISS spectrum was generated from the mean of all 475 spectra and fit using forward modelling. In forward modelling, observed spectra are fit with synthetic spectra from model atmosphere grids, each dimension of the grid corresponding to some fundamental parameter space, e.g. effective temperature ($T_{\rm eff}$), surface gravity (log $g$) and metallicity ([M/H])~\citep{Forward_2020}. We used five model grids: \texttt{ATMO} \citep{2015_ATMO}, \texttt{PHOENIX} \citep{2013_PHOENIX} and \texttt{Sonora Bobcat} \citep{2021_Sonora}, which use cloudless models, and \texttt{BTSettl(CIFIST)} \citep{2012_BT-settl} and \texttt{Sonora Diamondback} \citep{2024_Sonora}, which incorporate dust/clouds. The latter includes the sedimentation-efficiency parameter $f_{\rm sed}$, which describes the vertical distribution and particle size of the clouds. To access the grids, we used the python package \texttt{gollum} (v. 0.3.0), modified to support the \texttt{ATMO}, \texttt{BTSettl(CIFIST)} and \texttt{Sonora Diamondback} grids. 

We first applied rotational and instrumental broadening to match the resolution of the data. The fluxes of the degraded model spectra were then scaled to that of the average spectrum using a scaling factor $k_{\rm fit}$, determined using the same method as \cite{2023_Sanghi}:
\begin{equation} \label{sanghi}
    k_{\rm fit} = \frac{\Sigma(\bar{F}(\lambda)P(\lambda)/\sigma_{\bar{F}})}{\Sigma(\bar{F}(\lambda)^2/\sigma_{\bar{F}})}
\end{equation}

 where $\bar{F}(\lambda)$ and $P(\lambda)$ are the fluxes of the mean spectrum and degraded model spectra respectively, and $\sigma_{\bar{F}}$ is the uncertainty in the mean spectrum flux.

Following this, we performed a grid search for the minimum $\chi^2$ between the scaled model and average spectrum to produce the best fit spectrum for each model grid, as shown in Figure~\ref{fig:all}. 

Whilst each model grid produces poor residuals to our average spectrum, these fits are still informative. We found that for models allowing variable metallicity (\texttt{PHOENIX}, \texttt{Sonora Bobcat} and \texttt{Sonora Diamondback}), the best fits still preferred solar metallicity. However, there is a clear divide between the grids, with the non-dusty models preferring low surface gravities whilst \texttt{Sonora Diamondback} prefers a very high surface gravity and an $f_{sed} = 1$. \texttt{BTSettl(CIFIST)}, while incorporating dust, also prefers low surface gravity, but it has been suggested (e.g. \citealt{2020_Petrus, 2024_Hurt}) that not enough dust opacity is included within these models for M/L transition objects, potentially explaining this discrepancy.

The models lacking in dust prefer lower gravities because lower surface gravity leads to a reduced depth in molecular bands such as the H$_2$O band separating the JHK spectral windows. This produces a better fit to the data. These lower gravity fits align with previous literature~\citep{Berdyugina_2017}, which suggests that LSR~J1835 is very young. However, we find such youth unlikely. 

We do not see the "triangular" shape of the $H$-band peak/wide H$_2$O bands characteristic of lower gravities (e.g. \citealt{triangle_2013}). Furthermore, no lithium absorption feature at 6708 \AA~has been detected within its atmosphere ~\citep{2009LithiumLSR}, implying that it must be older than the lithium depletion timescale and have a mass greater than the threshold limit 65 $M_{\rm Jup}$~\citep{2023Kao}. \texttt{BANYAN $\Sigma$} analysis of the space velocity suggests candidate membership of the Corona of Ursa Major Association (CUMA)~\citep{Gagne_2018, 2026Gagne}, with an average age of 414 Myr~\citep{2015Jones}. Although CUMA has not yet been independently confirmed as a moving group, backtracking of the orbit using the Python package \texttt{galpy} \citep{galpy} shows consistency with CUMA being a plausible origin for LSR~J1835.  Combined, these factors indicate that it is likely LSR~J1835 is an older, higher surface gravity UCD.

The \texttt{Sonora Diamondback} grid prefers a more physical high surface gravity. However, it requires $f_{\rm{sed}} = 1$, which represents extremely thick clouds with essentially no sedimentation~\citep{2024_Sonora}. This dust has a similar impact on the emergent model spectrum to that of the low gravity in the other models, compressing the heights of the bands, but via an increase in opacity, reducing the emergent flux. Such a large quantity of dust, however, is probably unrealistic at these warm temperatures.

We therefore conclude that model grid spectra alone are unable to reproduce the average spectrum. Instead, we argue that the unrealistically low gravities, improbable quantities of dust and overall poor residuals are all consequences of using single atmosphere models to fit a heterogeneous surface. To therefore fit the average spectrum, we use a similar method to \cite{Hallinan_2015}, who suggested that a blend of an M-dwarf spectrum with a surface feature can be used to model the emission from LSR~J1835. This is explored further in section~\ref{Hallinan var sect}.

\subsubsection {Spectral Signatures of Aurora}

\begin{figure}
 \centering
 \includegraphics[width=\columnwidth]{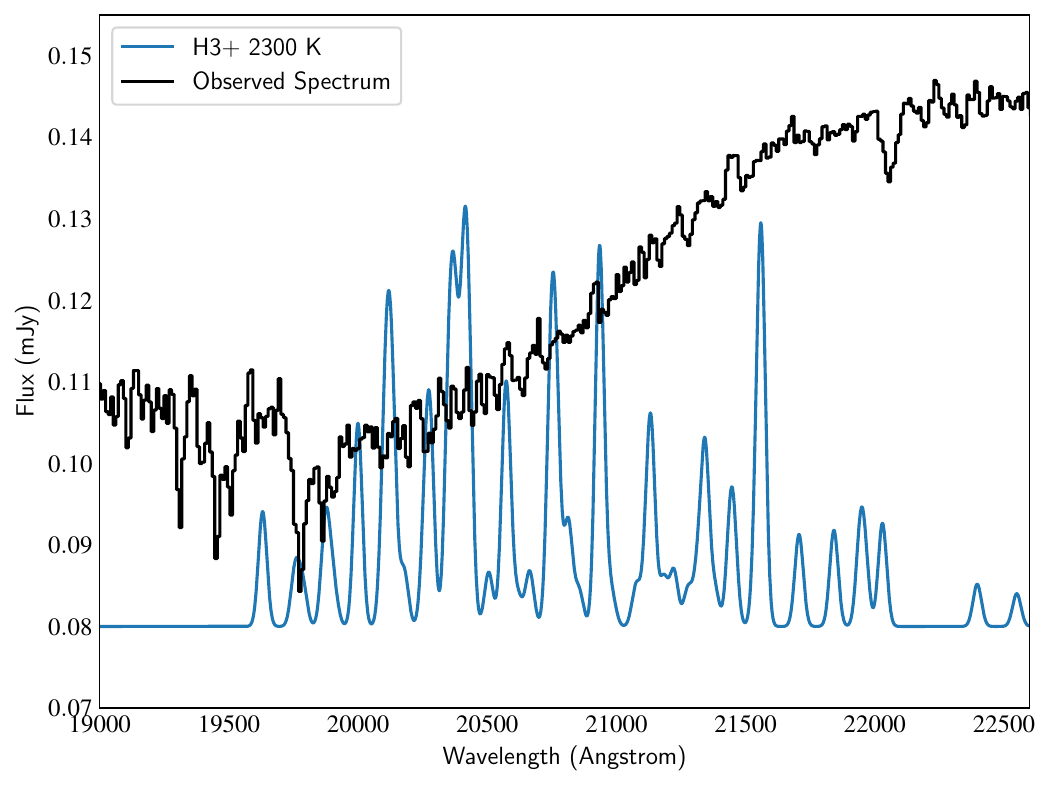}
 \caption{ The average spectrum of LSR~J1835+3259 (orange) is shown together with the expected emission spectrum from H$_3^+$ at 2300 K. See text for details of the computation of the H$_3^+$ emission spectrum. No features are visible at wavelengths corresponding to the expected emission lines from H$_3^+$.}  
 \label{fig:h3plus_2300}
\end{figure}

\begin{figure}
 \centering
 \includegraphics[width=\columnwidth]{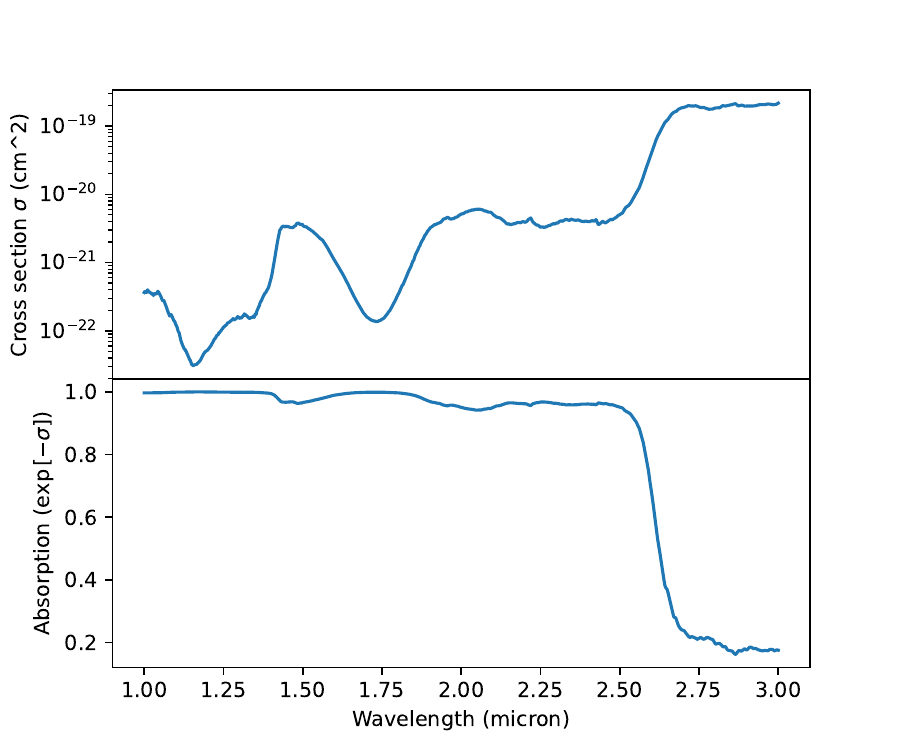}
 \caption{ \textbf{Top:} Cross sections  and \textbf{Bottom:} resulting absorption spectrum  from hydronium at 1500 K. See text for details of how the cross sections were computed.}  
 \label{fig:hydronium}
\end{figure}

One signature of auroral phenomena in the Solar System is the collisional ionization of molecular hydrogen and production of H$_3^+$. Spectral lines from the rotational-vibrational modes of H$_3^+$ are one of the strongest signatures of aurora from the Jovian ionosphere \citep{Maillard-Miller2011}. Searches for H$_3^+$ emission in auroral brown dwarfs have not yielded any detections to date \citep{Gibbs_Fitzgerald2022, Pineda_2024}.

Figure~\ref{fig:h3plus_2300} shows the average spectrum of LSR~J1835 in the region of the 2$\nu_2$ overtone band of H$_3^+$ where we expect potentially strong emission features. The expected H$_3^+$ spectrum is taken from the intensity calculator of the H$_3^+$ Resource Center \footnote{http://h3plus.uiuc.edu/}. The calculator uses input rotational and vibrational temperatures and a computed line list to construct the intensity spectra following~\cite{1996Neale}. We equate both temperatures and set them equal to the effective temperature of $\sim2300$\, K. No signs of H$_3^+$ emission are seen, consistent with previous studies.

\cite{Pineda_2024} suggest that H$_3^+$ emission may not be expected in auroral brown dwarfs. Auroral brown dwarfs have higher surface gravities than Jupiter, and possibly higher energy electron beams. Both effects drive the production of H$_3^+$ to deeper pressures, where the species may be destroyed before it can radiatively de-excite. \cite{Pineda_2024} suggest that destruction of H$_3^+$ via H$_2$O may lead to the production of hydronium (see also \citealt{Helling_Rimmer_2019}).

We produced Doppler-broadened cross-sections of hydronium using the line list of \cite{Yurchenko_2020} for an effective temperature of 1500 K, which is the highest temperature for which the line list is complete. Cross-sections were produced using \texttt{EXOCROSS} \citep{Yurchenk_2018}. The cross-sections are shown in Figure~\ref{fig:hydronium}. The clearest sign of hydronium would be an absorption band or emission bump at the red end of our data, starting at $\sim 2.56$\,$\mu$m. No such signature is seen in our data.

\subsection{Dynamic Spectra/PCA} \label{sec: Dynamic spectrum/PCA}

\begin{figure*}
 \includegraphics[width=0.9\textwidth]{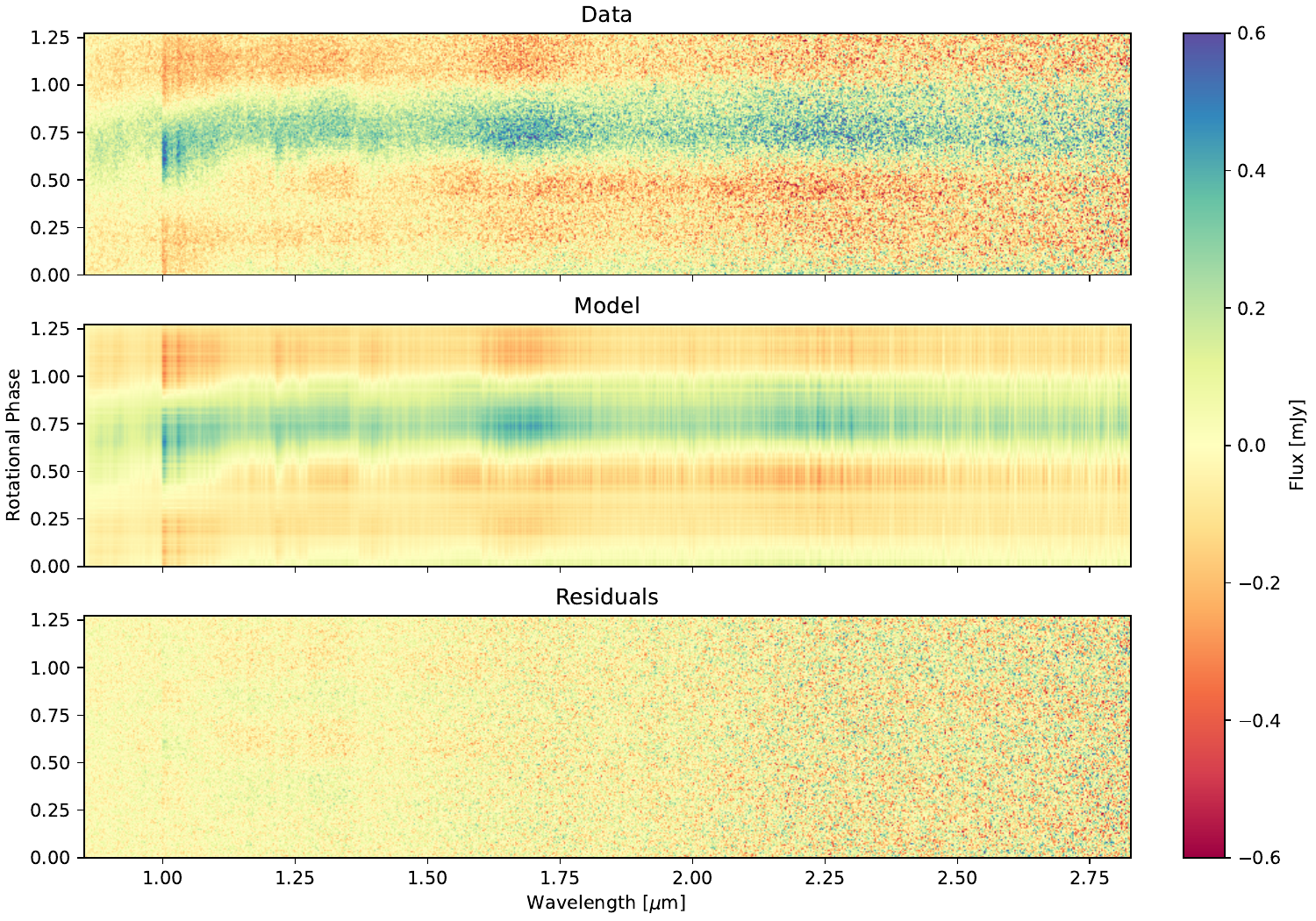}
 \caption{Dynamic spectra, showing the variability in flux as a function of wavelength. The mean spectrum has been subtracted off each spectrum. \textbf{Top}: Observed dynamic spectra. \textbf{Middle}: PCA reconstruction of the dynamic spectrum with two components. \textbf{Bottom}: Residuals between observed and PCA modelled dynamic spectra.}
 \label{fig:dynamic}
\end{figure*}

In order to better inform the blended spectral modelling in Section~\ref{Hallinan var sect} we first examine the wavelength-dependent variability present in the NIRISS time series. The variability in our NIRISS data is visualised via a dynamic spectrum in the top panel of Figure~\ref{fig:dynamic}. Unlike "variability maps" shown in other literature (e.g. ~\citealt{2024Biller, 2025Nasedkin}), we have not divided all spectra by the mean spectrum but subtracted it to show the absolute, not relative, change in flux density over time as a function of wavelength. From this plot, we see a periodic variability in flux across the entire spectrum, consistent with LSR~J1835's rotation period. This variability in flux is small however, with maximum deviations of $\sim0.5$ per cent from the average spectrum. This is significantly smaller than the $\sim3$ per cent variations observed by~\cite{Hallinan_2015} in the KECK/LRIS dataset (between $0.5-0.8~\mu$m), again aligning with the conclusion of Section~\ref{sec:Halpha} that LSR~J1835  may have been in a reduced activity state at the time of the NIRISS observations. This tiny variability is only detectable due to the exquisite precision and stability of \textit{JWST}, which provides 475 high-SNR spectra (median per-pixel SNR $\sim 267$), allowing coherent variability to be recovered despite its small amplitude.

To model this variability, we used principal component analysis (PCA), a dimensionality reduction technique that decomposes a dataset into orthogonal principal components, representing the dominant directions of variance, and associated scores that quantify the contribution of each component to the dataset~\citep{PCA_2016}. We used the \texttt{scikit-learn} python package to perform this decomposition with the \texttt{sklearn.decomposition.PCA} algorithm. The 475 NIRISS spectra were supplied in their original flux-density form, with the mean subtraction handled internally by \texttt{scikit-learn}. The PCA therefore describes the wavelength-dependent variability about the time-averaged spectrum. We set \texttt{whiten=False}, such that the PCA decomposition was performed without rescaling the transformed components to unit variance.

We also tested uncertainty-weighted PCA \citep{weighted_pca} using the Python package \texttt{wpca} and found negligibly different principal components and explained-variance fractions. The NIRISS SNR decreases toward longer wavelengths, with average values of approximately 450, 304, and 155 across the 0.85--1.49, 1.49--2.15, and 2.16--2.83~$\mu$m intervals, respectively. However, at a given wavelength the uncertainties remain approximately uniform across the time series, such that the relative data quality changes little between individual spectra. Accounting for the wavelength-dependent uncertainties therefore has little effect on the PCA results, and we retain the standard unweighted PCA for the analysis presented here.

We chose to use two components to model the variability, with the resulting PCA reconstruction of the dynamic spectrum shown in the middle panel of Figure~\ref{fig:dynamic}. These two components account for a total explained variance of approximately 8 per cent, with component 1 contributing the majority ($\sim$7 per cent). The relatively small explained-variance fraction reflects the intrinsically weak variability of LSR~J1835 compared to the measurement noise, since the variance associated with the astrophysical signal scales with the square of its variability amplitude (e.g. \citealt{PCA_2016}). In contrast, any additional components beyond the first two each account for $\lesssim0.5$ per cent of the variance (see Appendix~\ref{ap.variance}) and exhibit very little spectral structure. The two-component PCA reconstruction yields a reduced $\chi^2_\nu$ of 1.09 for the residuals, indicating that the residual scatter is consistent with the observational uncertainties. Adding further components does not significantly improve the fit. We therefore adopt two components as a conservative representation of the detectable variability in the NIRISS spectra.

We additionally explored the effect of temporal and wavelength binning prior to PCA. While binning increases the explained variance of the leading principal components by reducing the contribution of small-scale fluctuations, the dominant wavelength-dependent variability morphology is preserved. Since our subsequent analysis relies on retaining the spectral dependence of the variability, we retain the native NIRISS sampling throughout.

 Our PCA model is described by:
\begin{equation} \label{PCA}
    F(\lambda,t) - \bar{F}(\lambda) = C_1(t)\Delta_1(\lambda) + C_2(t)\Delta_2(\lambda) + \epsilon
\end{equation}

where $F(\lambda, t)$ is the flux density, $\bar{F}(\lambda)$ is the mean flux density, $C_i(t)$ are the time varying contributions of each component $\Delta_i(\lambda)$, and $\epsilon$ represents the residuals.

\begin{figure*}
 \includegraphics[width=\textwidth]{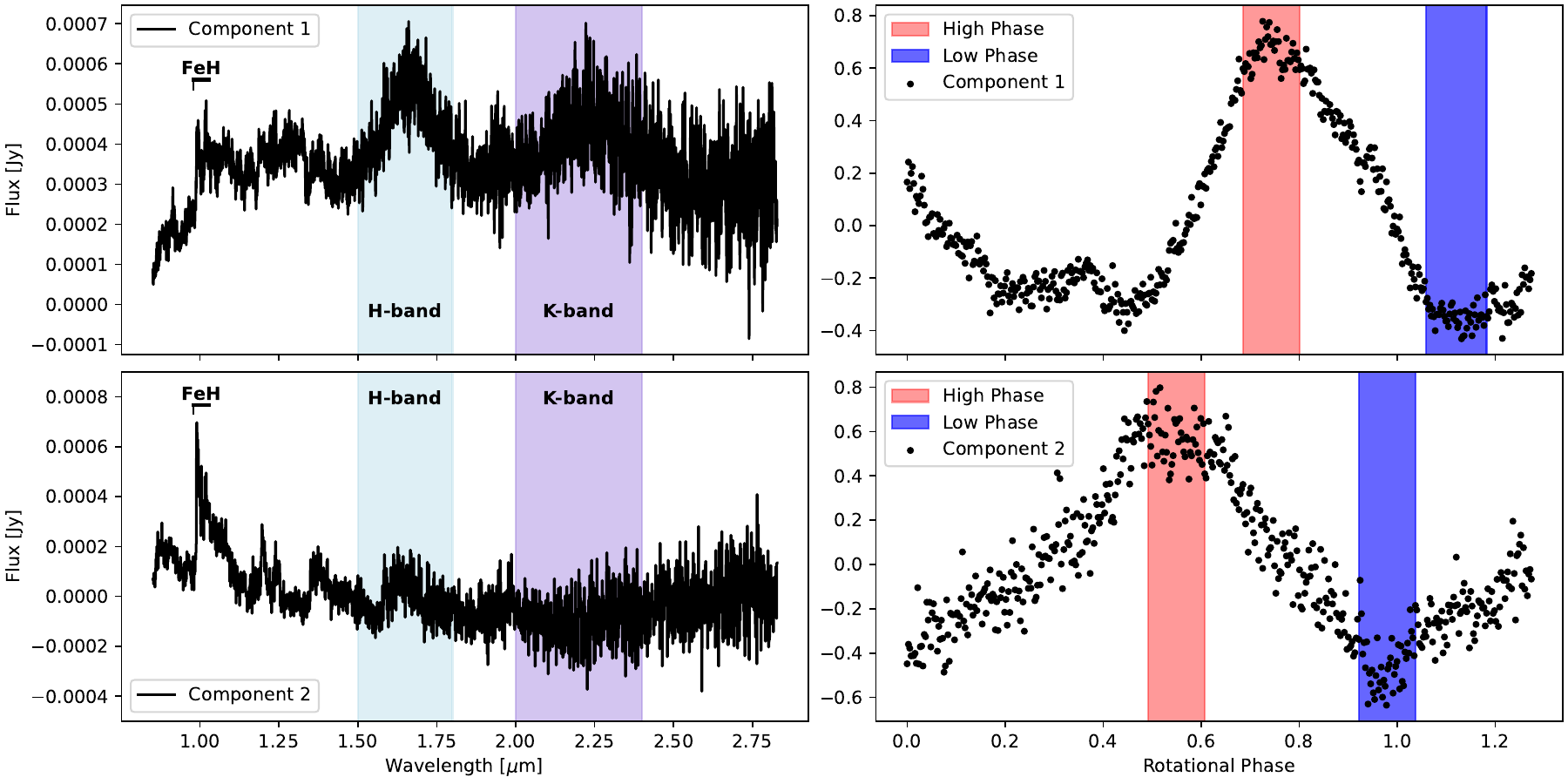}
 \caption{\textbf{Left}: the two PCA components used to model the variability of the NIRISS dataset. Component 2 is dominated by FeH (0.98-1.02~$\mu$m) variations whilst component 1 also has $H$ and $K$ -band variability. \textbf{Right}: Time-varying contributions of each component, with the phases corresponding to the high and low states indicated by the shaded regions. The high and low phases are used to generate difference spectra shown in Figure~\ref{fig:PCA_diff}.}
 \label{fig:pca}
\end{figure*}

The two components alongside their time-varying contributions are depicted in Figure~\ref{fig:pca}. In our data, the components represent the amplitude of variability as a function of wavelength. The time-varying contributions of both components exhibit quasi-sinusoidal modulation approximately on the rotational timescale, but are substantially phase-shifted. PCA components should not be interpreted as mapping directly onto discrete surface features (e.g.~\citealt{2011cowan, 2026schrader}). However, the distinct spectral signatures and phase behaviour of the two components suggest that they trace two distinct atmospheric structures or processes. Component 2 is dominated by FeH variations between 0.98--1.02~$\mu$m. Component 1 also shows this FeH variability alongside broadband modulation across the $H$ and $K$ -bands.

\subsection{Applying Hallinan 2015 Model} \label{Hallinan var sect}

Here we follow the simple spot model proposed by \cite{Hallinan_2015} (hereafter termed the Hallinan 2015 model) to interpret the variability seen in the two components. Generalising the model to two variable components, the flux density can be written as:
\begin{align}
     F(\lambda, t) &= \sum_{i=1}^{2}\alpha_i(t)S_i(\lambda) +  \Biggl [1-\sum_{i=1}^{2}\alpha_i(t) \Biggr]P(\lambda) \\
    & = \sum_{i=1}^{2}\alpha_i(t) [S_i(\lambda)-P(\lambda)]  + P(\lambda) \label{surface}
\end{align} 

where $S(\lambda)$ is the emission from a spot, $P(\lambda)$ the emission from the `immaculate' photosphere and $\alpha(t)$ the time-varying coverage fraction of a surface feature. The mean flux density is therefore:
\begin{align} \label{mean}
    \bar{F}(\lambda) = \sum_{i=1}^{2}\bar{\alpha_i} [S_i(\lambda)-P(\lambda)] + P(\lambda)
\end{align}

Subtracting the mean flux density from the flux and using equation~\ref{PCA}, we find:
\begin{align}
    F(\lambda,t) - \bar{F}(\lambda) &=  \sum_{i=1}^{2}[\alpha_i(t) - \bar{\alpha_i} ][S_i(\lambda)-P(\lambda)] \notag\\ 
    &\approx C_1(t)\Delta_1(\lambda) + C_2(t)\Delta_2(\lambda)
\end{align}

Rearranging and comparing variables, our two components should be reproducible via:
\begin{equation} \label{Hallinan Var}
    \Delta_i(\lambda) = \delta_i[S_i(\lambda)-P(\lambda)]
\end{equation}
where $\delta$ is some scaling factor.

\begin{figure}
 \includegraphics[width=\columnwidth]{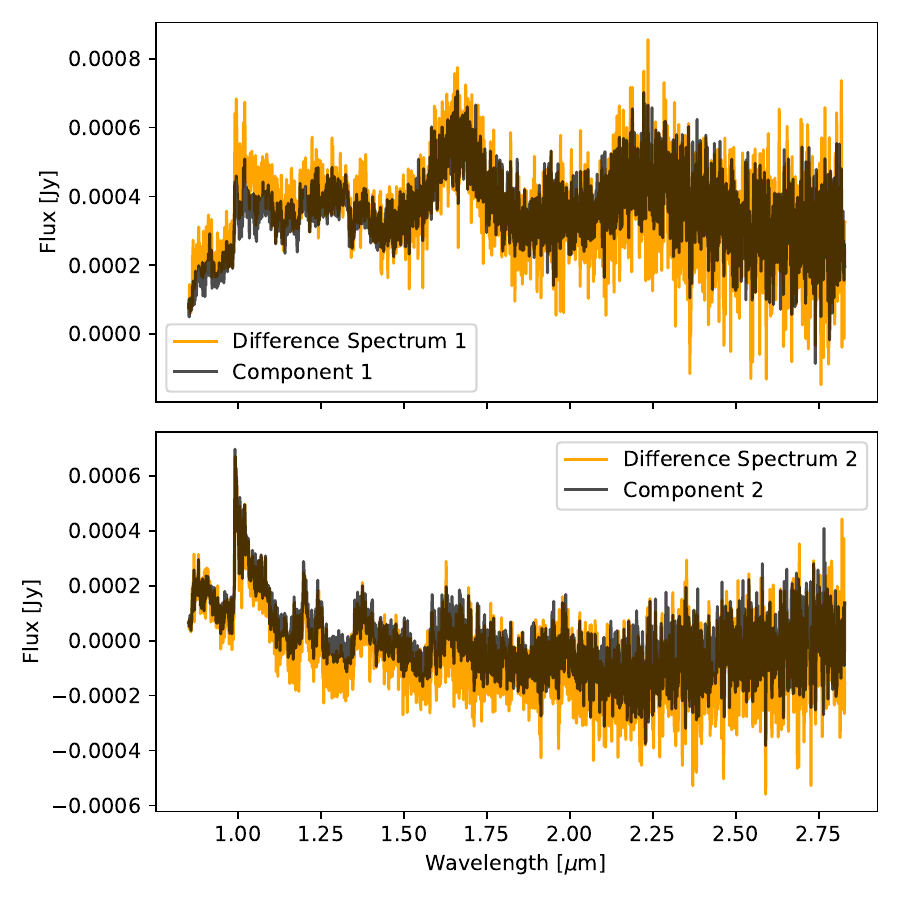}
 \caption{PCA components and corresponding difference spectra. The two are essentially equivalent representations of the amplitudes of variability, excluding deviations in FeH (0.98-1.02~$\mu$m) in component/difference spectrum 1. The scaling of the PCA components is described further in Appendix~\ref{ap.A}.}
 \label{fig:PCA_diff}
\end{figure}

However, the PCA components suffer from inherent degeneracies (see Appendix~\ref{ap.A}) and a lack of associated uncertainties. Therefore, similarly to \cite{Hallinan_2015}, we instead use `difference spectra' to represent the amplitude of variability in order to circumvent these issues. We first created two `high state' spectra by averaging all spectra between phases 0.685-0.801 and 0.492-0.607 for components 1 and 2 respectively. Similarly, `low state' spectra were created by averaging over 1.059-1.182 and 0.922-1.037. All high states and low states are shown by the shaded regions in Figure~\ref{fig:pca}. These ranges were chosen based on the PCA time-varying contributions such that the contamination from the other variability component would be approximately equal in both the high and low states. Subtracting the low states from the high states hence produced the difference spectra, which are shown in comparison to the PCA components in Figure~\ref{fig:PCA_diff}. The two representations of the amplitude of variability are essentially equivalent, though there is notably some FeH contamination in difference spectrum 1 (0.98-1.02~$\mu$m). However, as we believe the FeH modulation within both difference spectra is likely to be caused by a mechanism unaccounted for in our models (see below), we disregard this region in our following analysis.

Starspots are a common cause of variability in late M-dwarfs and therefore represent another possible explanation for the surface feature. Starspots in M-dwarfs have spectra similar to those of cooler M-dwarfs (e.g.~\citealt{starspots}). A two-temperature starspot model, represented as a blend of two stellar spectra with different temperatures, did not provide a good fit to the optical spectral variability presented by \cite{Hallinan_2015} (S. Littlefair, private communication). We similarly find that such models struggle to reproduce either the average spectrum or variability in our NIRISS spectra (see Appendix~\ref{ap. starspot}). We therefore disfavour starspots as the dominant cause of the variability.

\cite{Hallinan_2015} used an M8 template spectrum to represent the immaculate photosphere $P(\lambda)$ and determined that their single surface feature $S(\lambda)$ could be well approximated using a blackbody spectrum. They suggested this surface feature was an optically thick region at high altitudes, with negative hydrogen ions (H$^-$) as the dominant opacity source. This population of H$^-$'s would therefore be enhanced by the excess of free electrons as a result of the increased auroral electron number density.

It is immediately apparent from Figure~\ref{fig:PCA_diff}  (lower panel) that the Hallinan 2015 model cannot reproduce the variability in difference spectrum 2 as no mixture of an M-dwarf and blackbody spectrum would emulate this FeH dominated feature. We therefore interpret the  variability as arising from a separate physical mechanism, potentially the destruction or modification of FeH by the impacting auroral electron beam. Alternatively, this could result from more conventional atmospheric heterogeneity processes, such as cloud coverage and vertical transport. The FeH variability in difference spectrum 1 may also be influenced by the same mechanism as difference spectrum 2. Further progress in understanding this behaviour will likely require chemical network modelling that includes FeH formation and destruction pathways.

\subsubsection{Modelling the Average Spectrum and the Variability} \label{Niriss Hallinan}

To fit difference spectrum 1, we used the \texttt{BTSettl(CIFIST)} models with a log~$g$ = 5 and effective temperatures ranging from 2300-2600~K to model the emission of $P(\lambda)$. The emission from $S(\lambda)$ was modelled as a blackbody with temperature $T_{B}$. The scaling factor $\delta$ is a combination of two factors: $\alpha_{\rm diff}$, the difference in surface coverage between the high and low states, and $N = (R/d)^2$, with $R$ and $d$ being the radius and distance to LSR~J1835 respectively. $N$ is a dilution factor required to scale the fluxes of models $S(\lambda)$ and $P(\lambda)$ to the data as they are given in terms of flux crossing the unit surface area of the star. However, as the radius of LSR~J1835 is currently not well known, $\alpha_{\rm diff}$ and $N$ are degenerate factors. To partially break this degeneracy, and to obtain a more robust estimate of $N$, we therefore fit difference spectrum 1 simultaneously with the average spectrum using equation~\ref{mean}, neglecting the much smaller contribution from difference spectrum 2. 

Our free parameters $N$, $\bar{\alpha}$, $\alpha_{\rm diff}$ and $T_B$ were estimated using Markov-chain Monte Carlo (MCMC) analysis with the python package \texttt{emcee}~\citep{emcee}. We used flat priors for the surface coverages, requiring $0 <\bar\alpha < 1$ and $-1<\alpha_{\rm diff}<1$, and required $T_B$ to be positive. For $N$, we used a Gaussian prior with a mean of 1.83$\times 10^{-19}$ and standard deviation 0.1$\times 10^{-19}$. These values were calculated based on an estimated $R$, derived from \texttt{BTSettl} evolutionary models and observed luminosities, and the measured $d$~\citep{GaiaDr3} to LSR~J1835. We employed 200 walkers and MCMC chains consisting of 900 discarded burn-in steps followed by a production run of 10,000 steps, which ensured adequate convergence.

\begin{figure}
 \includegraphics[width=\columnwidth]{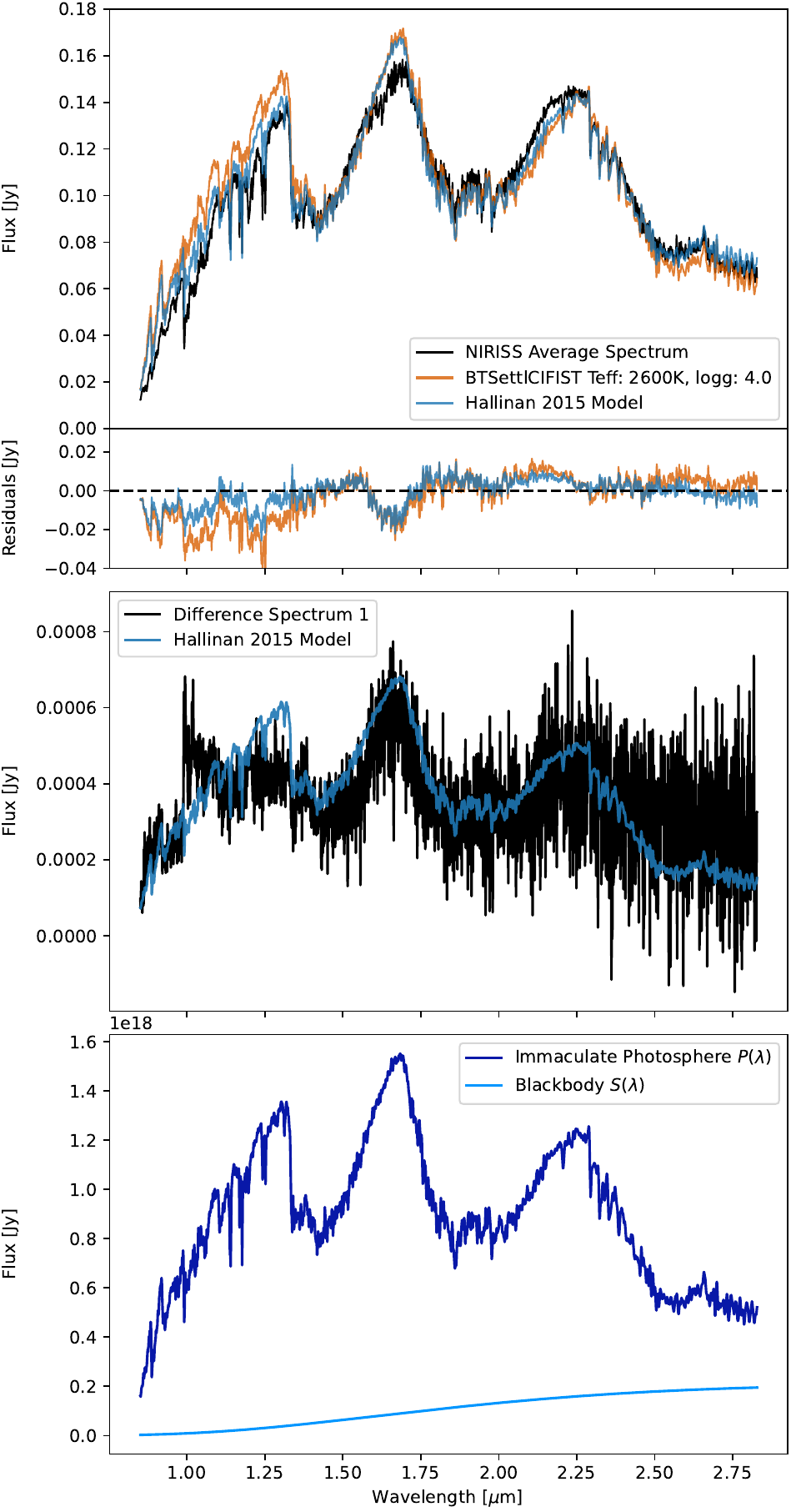}
 \caption{The best-fitting Hallinan et al. (2015) model applied to the NIRISS spectra. {\bf Top:} Fit to the time-averaged spectrum, including a comparison with the best-fitting single-atmosphere \texttt{BTSettl(CIFIST)} model ($T_{\rm eff}=2600$ K, $\log g=4.0$). The Hallinan model overestimates the flux at the $H$-band peak due to missing FeH opacity in the \texttt{BTSettl(CIFIST)} models. {\bf Middle:} Fit to difference spectrum 1. {\bf Bottom:} The components adopted for the Hallinan model: an immaculate photosphere $P(\lambda)$ (\texttt{BTSettl(CIFIST)}, $T_{\rm eff}=2600$ K, $\log g=5.0$) and a blackbody $S(\lambda)$ ($T_{\rm B}=1510$ K).}
 \label{fig:Hallinan Model}
\end{figure}

Figure~\ref{fig:Hallinan Model} shows our best fit model, with a $T_{\rm eff}$ = 2600~K, $T_B$ = 1510~K, $\bar{\alpha}$ = 0.500, $\alpha_{\rm diff}$ = -0.002, and N = 2.04 $\times 10^{-19}$. Overall, the model produces a reasonable fit to the average spectrum, providing an improvement on the fit achieved by the single spectrum models of section~\ref{section average}. This improvement is most obvious when comparing the molecular absorption lines of the $J$-band: the single \texttt{BTSettl(CIFIST)} fit of Figure~\ref{fig:all} consistently overestimates the flux, whereas the Hallinan 2015 model reproduces this region well. The $H$-band peak is not fit well, however, this is expected as \texttt{BTSettl(CIFIST)} models are known to be missing the 1.58~$\mu$m FeH transition~\citep{2024_Hurt}, therefore overestimating the flux. The Hallinan 2015 model also provides a broadly satisfactory fit to difference spectrum 1, although aside from the expected discrepancy around the 0.98–1.02~$\mu$m FeH feature, it also struggles to reproduce the 1.19–1.32~$\mu$m region, suggesting additional opacity or temperature structure not captured by the blackbody spot representation.

\subsubsection{Modelling the VOSA SED} \label{SED}

\begin{figure}
 \includegraphics[width=\columnwidth]{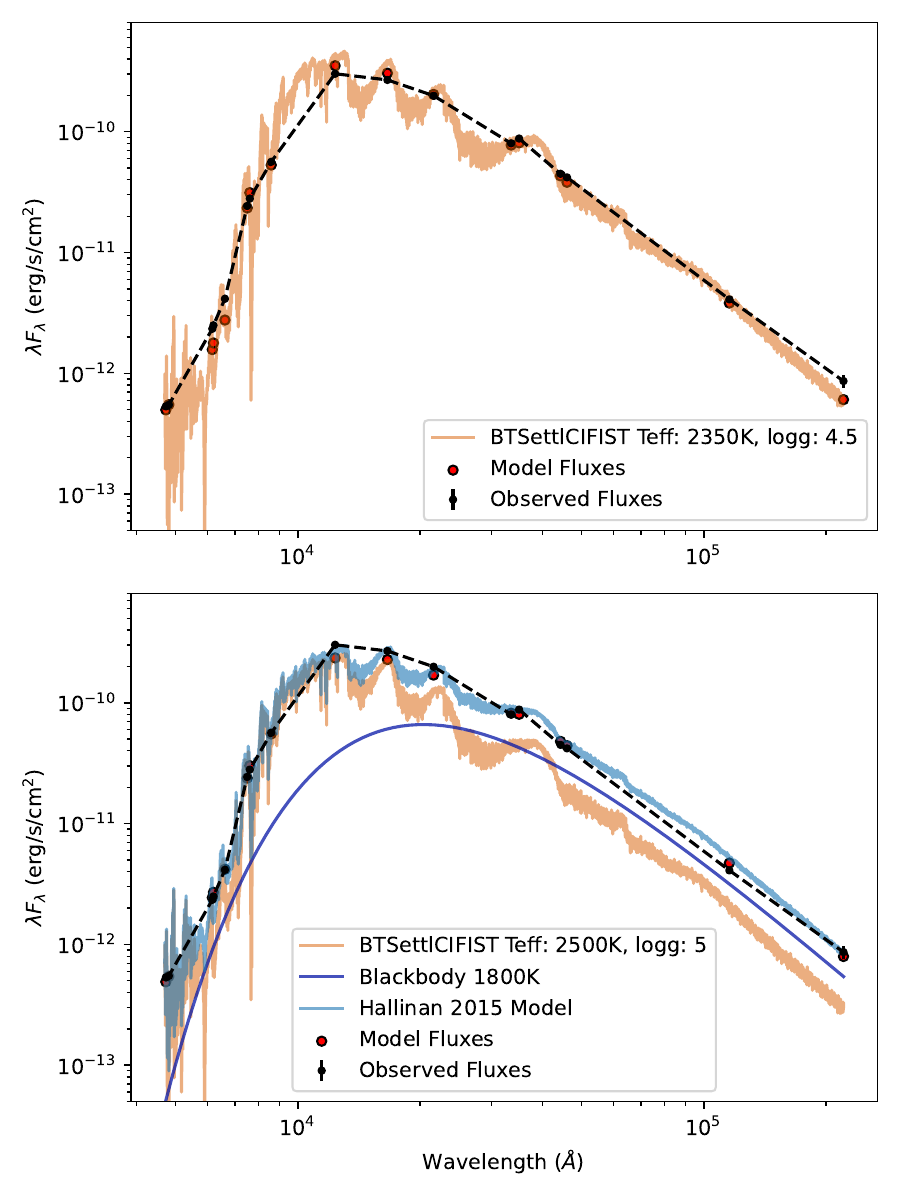}
 \caption{SED models of LSR~J1835, constructed and fit using VOSA. \textbf{Top}: Single M-dwarf spectrum fit ($T_{\rm eff} = 2350  \pm 25$, log $g$ = $4.5  \pm 0.25$). \textbf{Bottom}: the Hallinan 2015 model fit with M-dwarf ($T_{\rm eff} = 2500 \pm50$ K, log $g$ = $5  \pm0.25$) and blackbody ($T_B = 1800 \pm 25$ K) spectra. The blackbody component corresponds to a 63 per cent effective surface coverage.}
 \label{fig:SED}
\end{figure}

Alongside the difference and average spectra, we also performed fitting of the SED independently of the NIRISS data. The SED was constructed with VOSA: the Virtual Observatory SED Analyzer~\citep{Bayo2008}. This was done using photometry from the MOCA database \citep{2026Gagne}, excluding Gaia~\citep{GAIA_EDR3}, which appeared unreliable, and adding J-PLUS~\citep{JPLUS}, leading to 17 data points. To ensure a consistent comparison with the NIRISS-based analysis, we applied two models in VOSA: a single M-dwarf spectrum and a blend of M-dwarf and blackbody spectra (the Hallinan 2015 model). Both models used the \texttt{BTSettl(CIFIST)} model atmospheres to represent the M-dwarf spectrum. For the single model, 3 free parameters were fit: $T_{\rm eff}$, log $g$ and $N$. For the  Hallinan 2015 model, five parameters were fit: $T_{\rm eff}$, $\log g$, $T_{\rm B}$, and two dilution parameters, $N_1$ and $N_2$, which encode both the photospheric/feature flux scaling and the surface coverage fraction of the blackbody component. The best fits for both methods are shown in Figure~\ref{fig:SED}. 

The single-component model fit prefers a cooler $ T_{\rm eff} = 2350 \pm 25$\, K and lower gravity $\log g = 4.5 \pm 0.25$, whilst the Hallinan 2015 model gives $T_{\rm eff} = 2500 \pm 50$ K and higher gravity log$g = 5 \pm 0.25$, with a $T_{\rm B} = 1800 \pm 25$ K. The Hallinan 2015 model is therefore able to achieve a higher surface gravity, aligning with predictions of LSR~J1835 being an older object. 

Comparing the goodness-of-fit of the two models, the single-component model achieves a reduced $\chi^2_\nu  = 49.4$  whilst the Hallinan 2015 blended model has a $\chi^2_\nu = 17.3$. Thus, although neither achieves an ideal fit, the reduction in $\chi^2_\nu$ demonstrates that the Hallinan 2015 model offers a substantially better representation of the SED. This improvement is most evident in the optical between 6150 - 6700 \r{A}, where the single fit model overestimates the flux. 

Using the outputted luminosities and Stefan Boltzmann law to infer the  radius, the ratio of $N_1$ to $N_2$ implies a blackbody surface coverage of approximately 63 per cent. In Section~\ref{sec: perturbations}, we argue that the Hallinan 2015 model is not a physically self-consistent surface model. Whilst a blend of stellar atmosphere and blackbody is a good match to the SED, we do not believe any region of the surface actually looks like a blackbody. Consequently, the inferred 63 per cent coverage should not be interpreted as a direct measure of the physical surface area affected.

\subsection{Physical interpretation of the Hallinan 2015 model} \label{sec: perturbations}

\begin{figure*}
 \includegraphics[width=\textwidth]{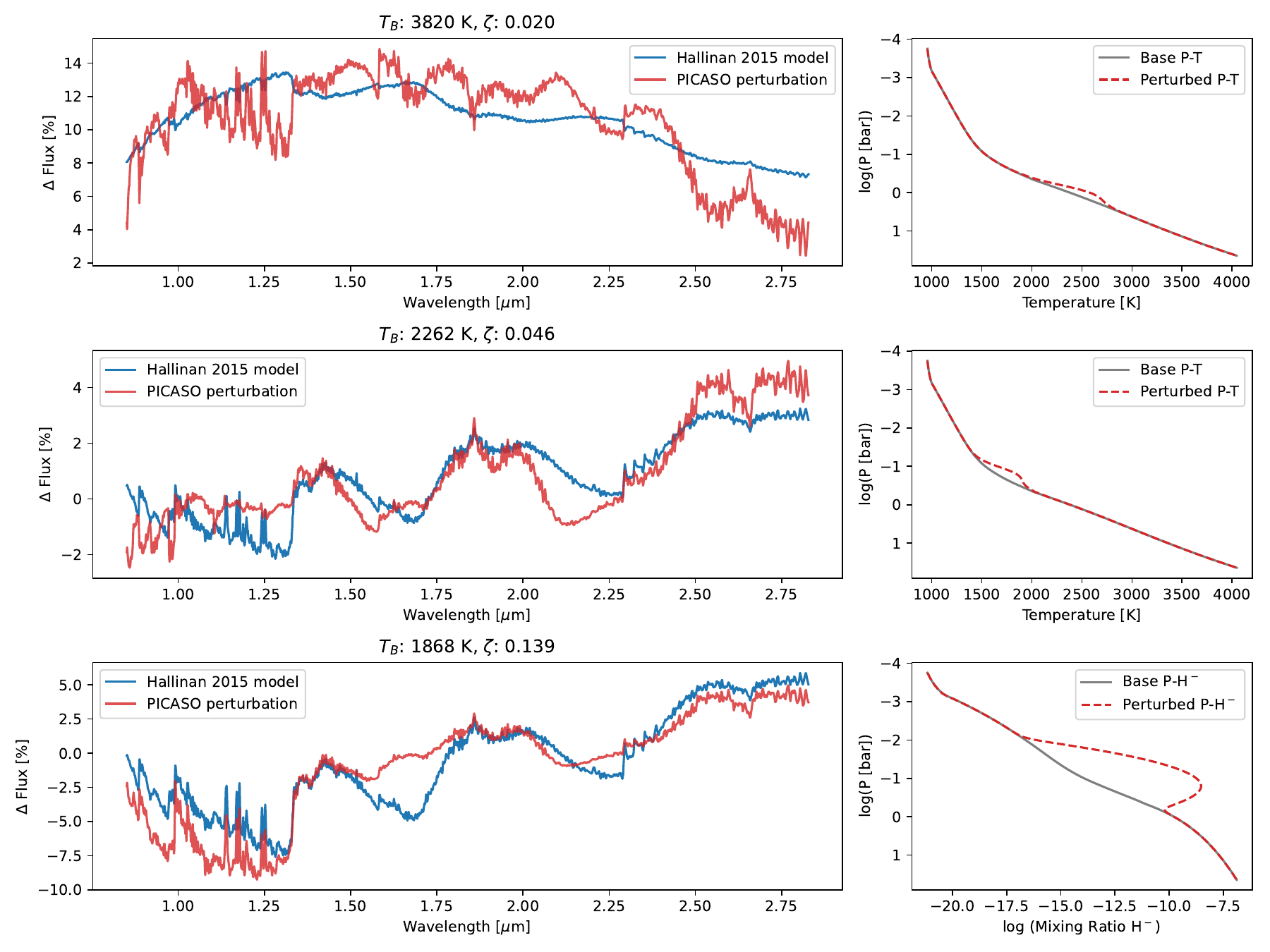}
 \caption{Emergent spectra following temperature and H$^-$ abundance perturbations and the corresponding best fit Hallinan 2015 models, both in the form of equation~\ref{eq: percentage change}. \textbf{Top}: Perturbation in temperature of 200~K at 1 bar. Right plot shows pressure-temperature profile, the solid line indicating the self-consistent \texttt{Sonora Diamondback} profile. \textbf{Middle}: Similar to top plots with perturbation of 200~K in temperature at  $10^{-0.8}$ bar. \textbf{Bottom}: Same 200~K perturbation in temperature as middle plot with additional perturbation in H$^-$ abundance at $10^{-0.8}$ bar. Right plot shows the mixing ratio profile of H$^-$.}
 \label{fig:perturbations}
\end{figure*}

\begin{figure}
 \includegraphics[width=\columnwidth]{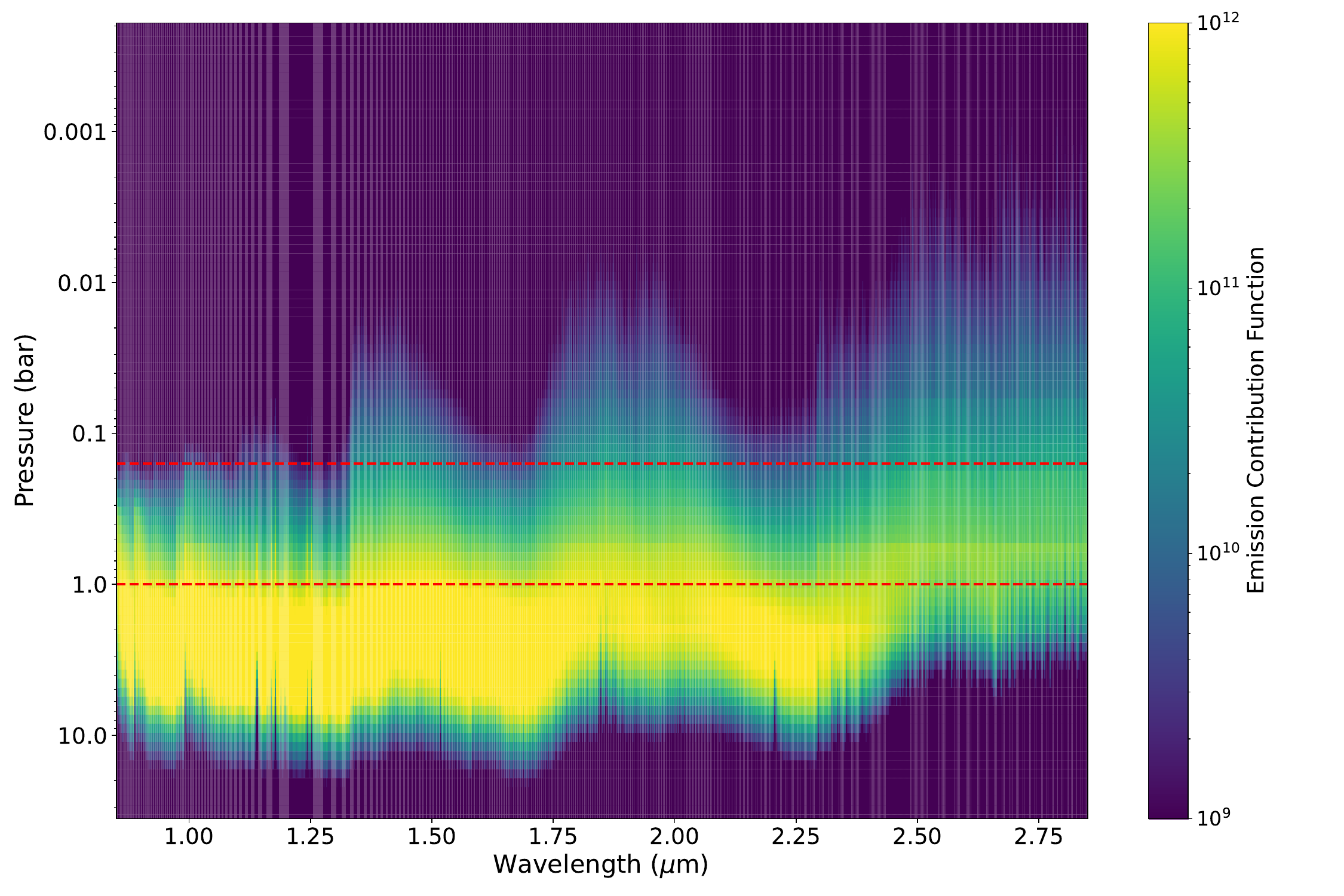}
 \caption{Emission contribution function of the \texttt{Sonora Diamondback} model used as the base atmosphere for the \texttt{picaso} perturbations, $T_{\rm eff}$ = 2400~K, log $g$ = 5. Dashed lines indicate $10^{-0.8}$ and 1 bar, the pressures where we add perturbations in Figure~\ref{fig:perturbations}. }
 \label{fig:contribution}
\end{figure}

In the Hallinan 2015 model, the blackbody surface feature is explained via an increase in H$^-$ opacity, which causes the dominant opacity source to be grey. This increase in opacity raises the height and lowers the temperature  of the photosphere, producing a cool blackbody spectrum. H$^-$ opacity undergoes a reduction in strength of several orders of magnitude redwards of 1.6~$\mu$m. However, our fitting of the average spectrum, difference spectrum 1  and SED all indicate that we require a blended photosphere and blackbody well beyond 1.6~$\mu$m. Therefore, an increase in H$^-$ opacity cannot be the sole explanation for this surface feature. An isothermal region of the atmosphere could also produce a blackbody spectrum, however this isothermal region would have to cover the full range of pressures probed by the NIRISS spectra and SED ($\sim$~0.01-10 bar). Such an extended isothermal region, covering approximately 60 per cent of the surface, seems unlikely a priori. 

This motivates investigation into the existence of more physically plausible models. In particular, we investigate if a mix of auroral heating and non-equilibrium chemistry can result in a surface feature whose emergent spectrum appears similar to the Hallinan 2015 model's mixture of a blackbody spectrum and M-dwarf photosphere. Auroral heating has already been invoked to explain emission from two UCDs (CWISEP J193518.59-154620.3; \citealt{2024Faherty} and SIMP J013656.5+093347.3; \citealt{2025McCarthy, 2025Nasedkin}), with energy deposited by precipitating electrons modifying the local temperature structure in the upper atmosphere. We also explore this heating working in tandem with auroral enhancement of the H$^-$ abundance.

A combination of heating and non-equilibrium chemistry would result in the spectrum of the auroral surface feature looking like a perturbed version of the immaculate photosphere. We therefore hypothesise that this perturbed spectrum looks like the blackbody and M-dwarf mix, explaining why the Hallinan 2015 model appears to work. That is:
\begin{align} \label{eq: perturbed}
    Q(\lambda) = N' [\zeta S(\lambda) + (1-\zeta)P(\lambda)]
\end{align} 

where $Q(\lambda)$ is the perturbed spectrum, $N'$ is some scaling factor, $\zeta$ is the fractional contribution of $S(\lambda)$, and $S(\lambda)$ and $P(\lambda)$ are as prior, with $S(\lambda)$ modelled by a blackbody spectrum. This formulation is effectively a two-column atmosphere model in which the emergent flux is a mixture of a perturbed (auroral) atmospheric column and an unperturbed photospheric column.

To test this hypothesis, we carried out perturbation simulations designed to mimic auroral heating and non-equilibrium chemistry. Our immaculate/base atmospheres were first generated from the \texttt{Sonora Diamondback}~\citep{2024_Sonora} pressure-temperature and chemical profiles using the \texttt{picaso} framework (v 3.0, \citealt{picaso_spectra, 2021_Sonora}). Currently, only the \texttt{Sonora} group of atmosphere models are available for use within the \texttt{picaso} framework, which is why we use them over the \texttt{BTSettl(CIFIST)}, which has a larger temperature range of models. The temperature and H$^-$ abundance profiles of the base atmosphere were then perturbed via a Gaussian in the log-pressure space, with the standard deviation $dp$ centred on some chosen pressure. If non-equilibrium chemistry was included, the temperature perturbation was applied first and the atmosphere recomputed in chemical equilibrium. The species abundance perturbation was then applied to this updated atmosphere. The resulting emission spectrum, $Q(\lambda)$, was then calculated for this perturbed atmosphere and convolved to the resolution of our NIRISS spectra. We modelled this $Q(\lambda)$ using equation~\ref{eq: perturbed} via least squares parameter estimation. 

Figure~\ref{fig:perturbations} shows three of our best fit models to $Q(\lambda)$ in terms of the percentage change from $P(\lambda)$, that is:
\begin{align} \label{eq: percentage change}
\frac{Q(\lambda)- P(\lambda)}{\bar{P}} \times 100
\end{align}
where $\bar{P}$ is the mean flux of the immaculate spectrum. For our base (immaculate) atmosphere, we used a 2400~K \texttt{Sonora Diamondback} model, log $g$ = 5 and no clouds. The top plot includes a 200~K temperature perturbation centred at 1 bar across a pressure width $dp = 0.2$, whilst the middle and bottom plots have this perturbation centred at $10^{-0.8}$ bar across the same pressure width. The bottom plot also includes an additional $3\times 10^{-9}$ mixing ratio amplitude perturbation in H$^-$ across the same pressures as the temperature perturbation. 

We then fit the spectra of these perturbed atmospheres with the Hallinan 2015 model to test whether it can reproduce the spectral response of a physically perturbed atmosphere. We find that the Hallinan 2015 model succeeds to varying degrees at mimicking the perturbed spectra. Blending a blackbody with an M-dwarf photosphere effectively weakens molecular band strengths, which is consistent with the effect of a steepened pressure–temperature profile induced by heating. A temperature perturbation at a deeper pressure (e.g. the top panel of Figure~\ref{fig:perturbations}) is therefore modelled by a Hallinan 2015 model with a hotter blackbody than one at lower pressures (e.g. middle panel of Figure~\ref{fig:perturbations}). 

A temperature perturbation at 1 bar corresponds to a pressure which is within the photosphere for many wavelengths (see Figure~\ref{fig:contribution}). It therefore produces a fairly uniform increase in flux as a function of wavelength. This fairly uniform flux increase produces a mild suppression of absorption bands across the spectrum, which is reasonably well recreated by a Hallinan 2015 model with a warm blackbody ($\sim 4000$ K).

By contrast, a perturbation at $10^{-0.8}$ bar lies above the main contribution-function peaks in Figure~\ref{fig:contribution}. It primarily enhances flux within the H$_2$O band-forming regions at $\sim$1.4, 1.9, and 2.5--2.75~$\mu$m, with a weaker effect near the CO overtone bandhead at $\sim$2.3~$\mu$m. The increase in flux at wavelengths corresponding to absorption bands produces a more marked reduction in the strengths of these absorption bands. This perturbation is well matched by a Hallinan 2015 model where the blackbody temperature is roughly equal to the effective temperature.

Modelling the observed spectra and variability of LSR~J1835 with the Hallinan 2015 model requires blackbodies with temperatures $\lesssim 1800$~K. The spectral variability introduced by such a model is shown in the bottom panel of Figure~\ref{fig:perturbations}. Compared to a Hallinan 2015 model with a warmer blackbody, this produces a larger flux change in the $J$-band and a rising trend towards the red end of the NIRISS spectrum. In order to recreate these spectral changes using a perturbed atmosphere, we required enhancements to H$^-$ abundance in addition to a temperature perturbation. The H$^-$ abundance enhancement suppresses flux bluewards of 1.6~$\mu$m, providing a better fit to the variations produced by a Hallinan 2015 model with a cooler blackbody.

These experiments show that the Hallinan 2015 model can approximate the spectral response of perturbed atmospheres. However, our perturbations assume identical pressure widths for temperature and H$^-$ enhancements; more sophisticated modelling allowing these perturbations to occupy different pressure layers would more accurately capture the influence of the auroral electron beam. Atmospheric retrievals of the NIRISS spectra by Wang et al (in preparation) will explore this.

We are currently unable to apply this perturbation model to the NIRISS spectra. This is due to the aforementioned limited temperature extent of the available atmosphere models that we can apply perturbations to using \texttt{picaso}. Previous modelling of the NIRISS average spectrum using the \texttt{BTSettlCIFIST} models (see section~\ref{Niriss Hallinan}) suggests we need a base atmosphere $T_{\rm eff} = 2600$~K, however, the \texttt{Sonora} group of atmosphere models have a maximum~$T_{\rm eff} = $ 2400~K. Auroral heating and non-equilibrium chemistry may therefore explain the observed variability in LSR~J1835, but we require higher temperature models and more detailed modelling (e.g. retrievals by Wang et al, in preparation) to fully verify this.

Recent work by~\cite{2026Zuckerman} computes the pressure depths associated with typical auroral electron beams for brown dwarf atmospheres, the suggested method by which these temperature and non-equilibrium chemistry perturbations could occur. They find that, for an atmosphere similar to LSR~J1835, an electron beam with typical energies of 100 keV produces peak ionisations at pressure levels of $\sim 10^{-4}$ bar. Higher energy beams penetrate more deeply in the atmosphere. Our observations suggest perturbations at pressure levels of 0.1--1 bar. For the perturbations we simulate in this work, we would require very high electron energies to probe these pressure regions (of order 1 MeV). Future modelling of these perturbations may therefore be useful in constraining the electron beam energies powering aurorae on UCDs like LSR~J1835.

\section{Conclusions} \label{sec: conclusions}

We have presented new \textit{JWST}/NIRISS time-series spectroscopy of the auroral UCD LSR~J1835, providing the most detailed near-IR view to date of its spectral variability. These observations extend the work of \cite{Hallinan_2015} into a new wavelength regime, enabling us to directly test proposed interpretations of the aurorally-driven variability. We summarise our key findings below:
\begin{itemize}
    \item The variability in H$\alpha$ emission has decreased since 2012, suggesting LSR~J1835 may have been in a lower activity state at the time of the NIRISS observations.

    \item  No H$_3^+$ or hydronium signatures are seen within the NIRISS data, placing constraints on the ion chemistry expected from auroral processes in UCD atmospheres.

    \item PCA reveals that two components are required to reproduce the observed spectral variability. The second spectral component is dominated by FeH variations between 0.98--1.02~$\mu$m whilst the first includes additional variability in the $H$ and $K$ -bands. This FeH modulation may be aurorally driven but requires further work, such as chemical network modelling, to fully understand its origin. 

    \item Our forward-modelling demonstrates that model atmosphere grids fail to successfully reproduce the time-averaged NIRISS spectrum, suggesting a heterogeneous surface. Following the Hallinan 2015 model, we instead find that a two column model composed of an M-dwarf atmosphere blended with a blackbody component can fit the NIRISS average spectrum and the main component of the IR variability, as well as the broadband SED. 

    \item While the Hallinan 2015 model successfully reproduces the emission from LSR~J1835, it requires a cool blackbody covering a substantial percentage of the surface, making the model physically implausible. Further, the original H$^{-}$ interpretation cannot account for the full wavelength range, since H$^{-}$ opacity decreases by several orders of magnitude redwards of 1.6~$\mu$m, whilst the fitted blackbody contribution extends from optical to far-IR wavelengths.

    \item Using \texttt{picaso} to simulate auroral heating and non-equilibrium chemistry, we find that either or both can produce spectra that mimic a blend of unperturbed photosphere and a blackbody spectrum. This provides a more physically plausible explanation for the findings of \cite{Hallinan_2015} and the data presented in this paper. The apparent coolness of the blackbody mixture used to model the \textit{JWST} data and SED suggests that both auroral heating and non-equilibrium enhancement of H$^-$ are required to explain the spectrum and variability of LSR~J1835. However, we require hotter atmosphere models ($T_{\rm eff} \geq 2600$~K) in order to apply this perturbation model to the NIRISS data. This highlights the need for future high-temperature model grids and more sophisticated retrievals to fully assess the potential role of auroral heating in shaping the observed spectra.
\end{itemize}

Future simultaneous multi-wavelength observations of LSR~J1835, particularly coordinated optical, IR, and radio monitoring over multiple rotations, will be important for establishing the connection between auroral activity and the observed atmospheric variability. Extending this approach to auroral UCDs spanning a range of effective temperatures and auroral strengths would provide valuable constraints on the physical effects of auroral electron precipitation and help develop more realistic models of auroral heating and non-equilibrium chemistry in UCD atmospheres.

\section*{Acknowledgements}

We thank Merle Schrader for helpful discussion and valuable comments on the principal component analysis. EP acknowledges support from  UK Research and Innovation Science and Technology Facilities
Council [ST/Y509541/1]. FW and BB acknowledge support from UK Research and Innovation Science and Technology Facilities Council [ST/X001091/1]. This work is based in part on observations made with the NASA/ESA/CSA James Webb Space Telescope. The data were obtained from the Mikulski Archive for Space Telescopes at the Space Telescope Science Institute, which is operated by the Association of Universities for Research in Astronomy, Inc., under NASA contract NAS5-03127 for \textit{JWST}. These observations are associated with program \#1874. This work was partially funded by Space Telescope Science Institute under award \textit{JWST} 1874. This work is (partly) based on data obtained with MEGARA instrument~\citep{2018Gil, 2018Carrasco}, funded by European Regional Development Funds (ERDF), through Programa Operativo Canarias FEDER 2014-2020. This research made use of the Montreal Open Clusters and Associations (MOCA) database, operated at the Montr\'eal Plan\'etarium~\citep{2026Gagne}. This work is (partly) based on data from the CARMENES data archive at CAB (CSIC-INTA). The CARMENES archive is part of the Spanish Virtual Observatory project (http://svo.cab.inta-csic.es), funded by MCIN/AEI/10.13039/501100011033/ through grant PID2020-112949GB-I00.

%%%%%%%%%%%%%%%%%%%%%%%%%%%%%%%%%%%%%%%%%%%%%%%%%%
\section*{Data Availability}

All data can be made available upon
reasonable request.



\bibliographystyle{mnras}
\bibliography{example} % if your bibtex file is called example.bib

%\begin{thebibliography}{99}
%\end{thebibliography}

%%%%%%%%%%%%%%%%%%%%%%%%%%%%%%%%%%%%%%%%%%%%%%%%%%

%%%%%%%%%%%%%%%%% APPENDICES %%%%%%%%%%%%%%%%%%%%%

\appendix
\section{PCA Explained Variance} \label{ap.variance}
\begin{figure}
 \includegraphics[width=\columnwidth]{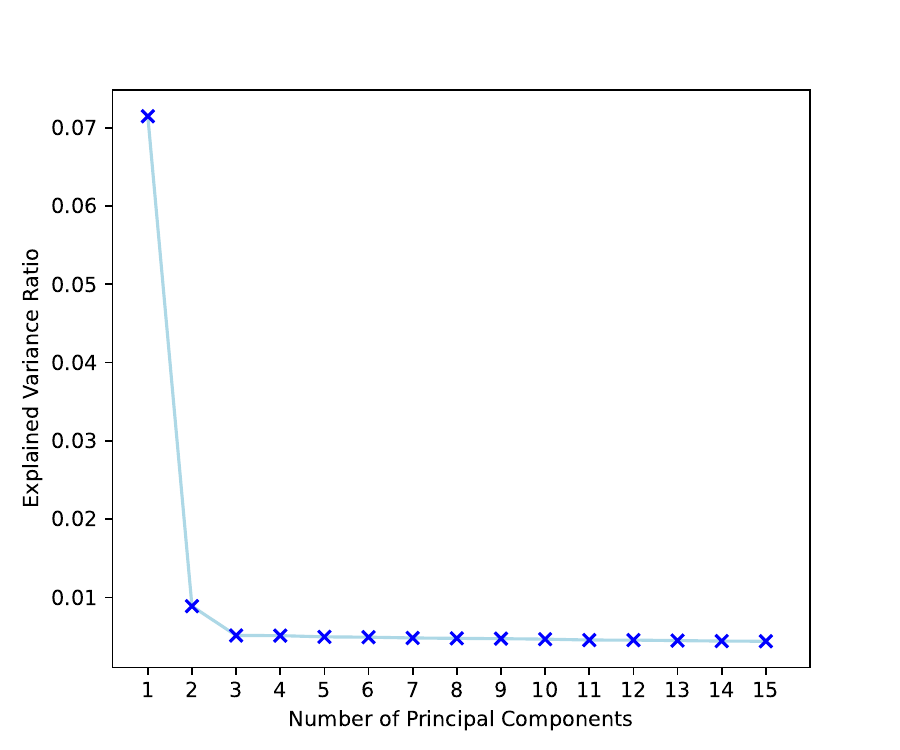}
 \caption{Individual explained variance for first 15 principal components, limited for display purposes. Only $\sim 8$ per cent of the variance is explained by the first two components, however, the remaining components individually contribute $\lesssim 0.5$ per cent to the explained variance.}
 \label{fig:explained}
\end{figure}

The explained variance of the PCA components is shown in Figure~\ref{fig:explained}. The individual components beyond the first two components each account for $\lesssim 0.5$ per cent and lack any significant spectral structure. We therefore only consider the first two components in our modelling of the variability.

\section{Scaling PCA Components} \label{ap.A}

The absolute change in flux from the time-averaged spectrum at time $t$, ignoring any errors, is:

\begin{align}
    F(\lambda,t) - \bar{F}(\lambda) =  C_1(t)\Delta_1(\lambda) + C_2(t)\Delta_2(\lambda)
\end{align}

Difference spectrum 1 is therefore equal to:
\begin{align}
    F_{1, \rm high} - F_{1, \rm low} =  [C_{1, \rm high} - C_{1, \rm low}]\Delta_1(\lambda) + \\
    [C_{2,\rm high}-C_{2,\rm low}]\Delta_2(\lambda)
\end{align}
where $F_{1, \rm high}$ and $F_{1, \rm low}$ are the high and low state spectra as described in Section~\ref{Hallinan var sect}. Combining all component 2 factors into a single term $x$:

\begin{align}
    x = [C_{2, \rm high} - C_{2, \rm low}]\Delta_2(\lambda)
\end{align}

Rearranging, we find:

\begin{align}
    F_{1, \rm high} - F_{1, \rm low} - x = [C_{1, \rm high}-C_{1, \rm low}]\Delta_1(\lambda)\\
     = \eta [C_{1,\rm high}-C_{1,\rm low}]\frac{\Delta_1(\lambda)}{\eta}
\end{align}

with constant $\eta$ included to show the degeneracy inherent to PCA. To scale our PCA component, we expect that our difference spectrum should be equal to component 1 divided by some constant, that is:
\begin{align}
    F_{1, \rm high} - F_{1, \rm low} - x =\frac{\Delta_1(\lambda)}{\eta}
\end{align}

Therefore, 
\begin{align}
    \eta = \frac{1}{C_{1,\rm high}-C_{1,\rm low}}
\end{align}

Figure~\ref{fig:PCA_diff} shows calculated difference spectra although note that this does not include the $x$ terms to fully remove contamination from the other component. We see that in the case of component 1, the two spectra are broadly equivalent, excluding wavelengths < 1.03~$\mu$m. 

\section{Starspot Model} \label{ap. starspot}

To investigate whether cool starspots could explain the observed variability, we modelled the surface of LSR J1835 as a combination of two atmosphere spectra with different effective temperatures and a common surface gravity. We tested three model grids: \texttt{BTSettl(CIFIST)}, \texttt{PHOENIX} and \texttt{Sonora Diamondback}. The relative contributions of the two spectra were allowed to vary, following the two-component formulation described in Section~\ref{Hallinan var sect}. We simultaneously fitted the time-averaged spectrum and difference spectrum 1. For each combination of atmosphere models, we used $\chi^2$ minimisation to optimise the mean surface coverage, change in surface coverage, and overall normalisation.

\begin{figure}
 \includegraphics[width=\columnwidth]{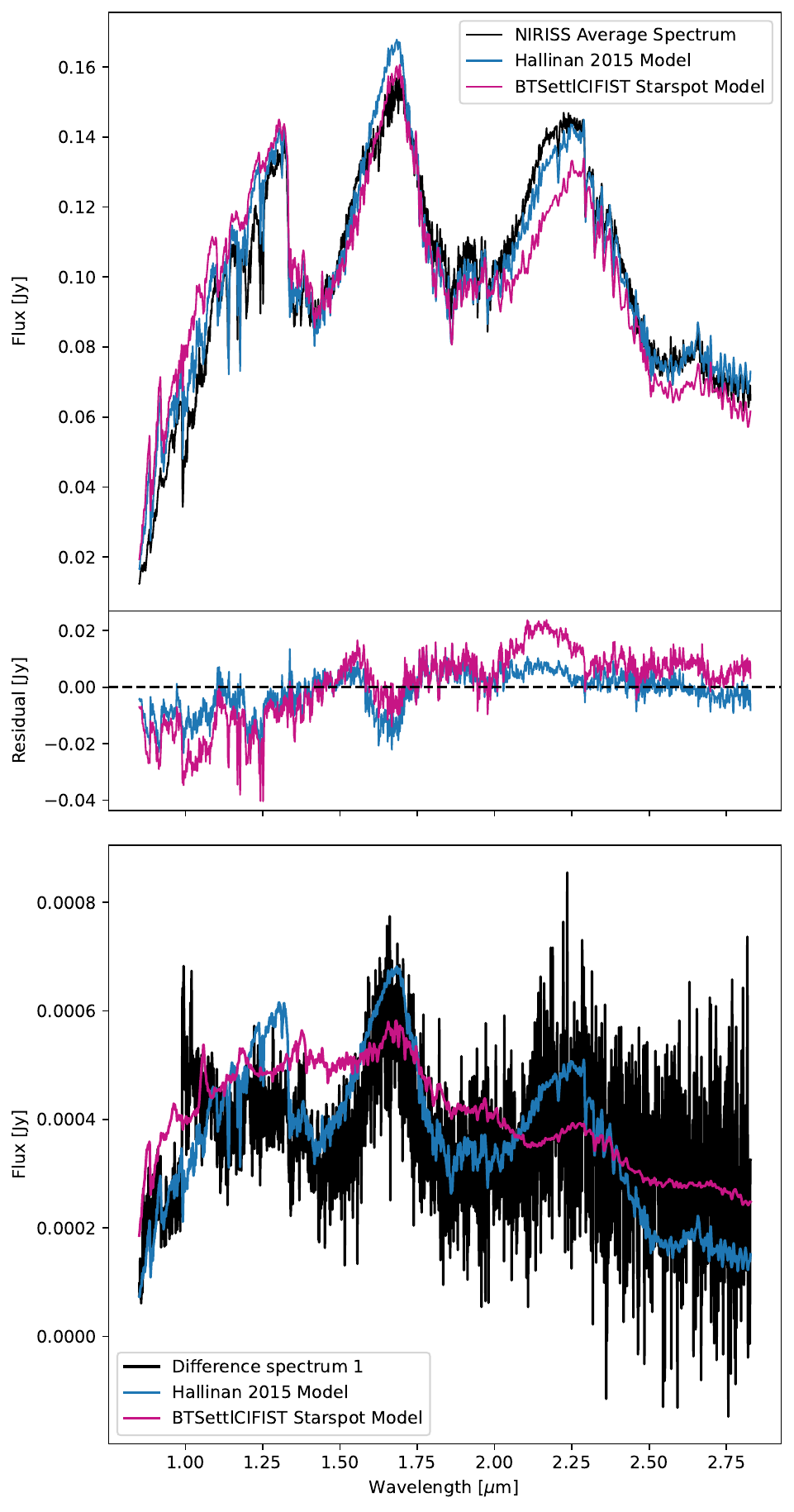}
 \caption{f Time-averaged spectrum (top), corresponding residuals (middle), and difference spectrum 1 (bottom) compared with the best-fitting Hallinan 2015 model and the two-temperature \texttt{BTSettl(CIFIST)} starspot model. The starspot model consists of a 2600 K cooler component and a 2700 K photosphere, both with $\log g = 3.5$. Although the starspot model broadly reproduces the variability beyond 2 $\mu$m, it struggles to reproduce the overall spectral morphology, particularly at shorter wavelengths where it overestimates the variability around 1.4 $\mu$m.}
 \label{fig:starspot}
\end{figure}

Of the three atmosphere grids, \texttt{BTSettl(CIFIST)} provided the best fit, with a 2600 K cooler component and a 2700 K photosphere, both with $\log g = 3.5$ (see Figure~\ref{fig:starspot}). Although the temperature contrast of 100 K is plausible for cool starspots, the preferred surface gravity is substantially lower than expected for LSR J1835, as discussed in Section~\ref{section average}. The model also fails to reproduce the time-averaged spectrum in detail, with systematic residuals across several wavelength regions, particularly in the $J$ and $K$ -bands, where it overestimates and underestimates the observed flux, respectively. Furthermore, the model struggles to reproduce the wavelength dependence of the observed variability. While it broadly reproduces the variability at longer wavelengths ($\gtrsim 2~\mu$m), it does not adequately reproduce the spectral structure at shorter wavelengths. In particular, the model overestimates the variability around 1.4 $\mu$m, failing to reproduce the pronounced contrast between the $J$ and $H$ -bands.

These discrepancies suggest that a simple two-temperature starspot model cannot adequately reproduce the broadband variability represented by difference spectrum 1. We therefore disfavour conventional cool starspots as the dominant source of this variability.

%%%%%%%%%%%%%%%%%%%%%%%%%%%%%%%%%%%%%%%%%%%%%%%%%%

% Don't change these lines
\bsp	% typesetting comment
\label{lastpage}
\end{document}